%% file: main.tex
\documentclass[preprint,12pt,authoryear]{elsarticle}
\usepackage{amsmath}
\usepackage{amssymb}
\usepackage[dvips]{epsfig}
\usepackage{dcolumn}
\usepackage{enumerate}
\usepackage{hhline}
\usepackage{dsfont}
\usepackage{afterpage}
\usepackage{arydshln}
\usepackage{graphicx}
\usepackage{color}
\usepackage[usenames,dvipsnames]{xcolor}
\usepackage{rotating}
\usepackage{xr}
\usepackage{algorithmicx}
\usepackage[noend]{algpseudocode}
\usepackage{algorithm}
\usepackage{diagbox}
\usepackage{graphicx}
\usepackage{wrapfig}
\usepackage{lscape}
\usepackage{fontenc}
\usepackage{setspace}
\usepackage{bm}
\usepackage{slashbox}
\usepackage{lscape}
\usepackage{breakurl} 
\usepackage{multirow}
\usepackage{eurosym}

\input{defs.tex}

\begin{document}
\input{title}
\input{sec1}
\input{sec2}
\input{sec3}
\input{sec4}
\input{sec5}
\input{sec6}
\input{sec7}

\input{append}
\bibliographystyle{elsarticle-harv}
\bibliography{references}
\end{document}

%% file: defs.tex
\def \avec {\text{\boldmath$a$}}

\def \evec {\text{\boldmath$e$}}

\def \qvec {\text{\boldmath$q$}}

\def \uvec {\text{\boldmath$u$}}    
\def \vvec {\text{\boldmath$v$}}    
    
\def \xvec {\text{\boldmath$x$}}    
\def \yvec {\text{\boldmath$y$}}    
\def \zvec {\text{\boldmath$z$}}

\def \betavec         {\text{\boldmath$\beta$}}

\def \deltavec        {\text{\boldmath$\delta$}}
\def \epsilonvec      {\text{\boldmath$\epsilon$}}

\def \zetavec         {\text{\boldmath$\zeta$}}
\def \etavec          {\text{\boldmath$\eta$}}
\def \thetavec        {\text{\boldmath$\theta$}}

\def \lambdavec       {\text{\boldmath$\lambda$}}
\def \muvec           {\text{\boldmath$\mu$}}

\def \rhovec          {\text{\boldmath$\rho$}}

%% file: title.tex
\begin{frontmatter}


\title{Implicit Copulas: An Overview}
\author[1]{Michael Stanley Smith}
\ead{mikes70au@gmail.com}
\affiliation[1]{organization={Melbourne Business School, University of Melbourne},
	addressline={200 Leicester Street},
	city={Carlton},
	postcode={3053},
	country={Australia}}
\date{}

\begin{abstract}
Implicit copulas are the most common copula choice for modeling
dependence in high dimensions. 
This broad class of copulas is introduced and surveyed, including
elliptical copulas, skew $t$ copulas, factor copulas, time series copulas and regression copulas. 
The common auxiliary representation of implicit copulas is outlined,
and how this makes them both scalable and tractable for statistical modeling. 
Issues such as parameter identification, extended likelihoods
for discrete or mixed data, parsimony in high dimensions, and simulation from the copula
model are considered.
Bayesian approaches to estimate the copula parameters, and predict from an
implicit copula
model, are outlined. 
Particular attention is given 
to implicit copula processes constructed from time series and regression models, which is 
at the forefront of current research. 
Two econometric applications---one from macroeconomic time
series and the other from financial asset pricing---illustrate the advantages of 
implicit copula models.	
%
%
%
\end{abstract}

\begin{keyword}
	copula process \sep factor copula \sep inversion copula \sep regression copula  \sep skew $t$ copula,
	time series copula
\end{keyword}



%
%
%
%
 
\end{frontmatter}

%% file: sec1.tex

\section{Introduction}\label{sec:intro}
Copulas are widely used to specify multivariate distributions 
for the statistical modeling of data. 
Fields where copula models have had a significant impact include 
(but are not limited to) actuarial science~\citep{frees1998}, finance~\citep{cherubini2004,McNFreEmb2005,patton2006}, hydrology~\citep{favre2004,genest2007meta}, climatology~\citep{schoelzel2008},
transportation~\citep{bhat2009,smithbike2011} and marketing~\citep{danaher2011,park2012}. 
Copula models are popular because they simplify the specification of a distribution, allowing
the marginals to
be modeled arbitrarily, and then combined using a copula function. In practice, 
a major challenge is the selection and estimation of
a copula function that captures the dependence structure well and is tractable. 
One choice are
``implicit copulas'', which are copulas constructed from existing multivariate distributions by the
inversion of Sklar's theorem as in \citet[p.51]{nelsen06}.
This is a large and flexible family of copulas, which
share an auxiliary representation that makes estimation tractable in high dimensions. 
Thus, they are suitable for modeling the large datasets that arise in many 
modern applications.
The objective of this paper
is to introduce and survey implicit copulas and their use in statistical modeling in an accessible manner.

Implicit copulas
have a long history with key developments spread across multiple fields,
including actuarial studies, econometrics, operations research,
probability and statistics.
Yet while there are many excellent existing monographs and surveys on copulas and 
copula models (see~\cite{genest1986,joe97,McNFreEmb2005,nelsen06,genest2007,jaworski2010,patton2012,nikoloulopoulos2013,joe2014dependence} and~\cite{durante2015} for prominent examples) 
there does not appear
to be a dedicated survey or overview on this important class of copulas. 
This paper aims to fill this gap and provides an overview that stresses common
features of the implicit copula family, likelihood-based estimation, and the usefulness
of implicit copulas in statistical 
modeling. Particular focus is given to recent developments on 
implicit copula processes for regression
and time series data, along with Bayesian inference that extends the 
earlier overview by~\cite{smithbcop2013} to these copula processes.

Two econometric applications illustrate
the use of implicit copula models with non-Gaussian data. 
The first is a time-varying heteroscedastic time series model for U.S. inflation 
between 1954:Q1 and 2020:Q2. The implicit copula is a copula process 
constructed from
a nonlinear state space model as in~\cite{smithman2018}. It is a 
``time series copula'' that captures serial dependence.
The second application is a five factor asset pricing regression model \citep{fama2015five}
with an asymmetric Laplace marginal distribution for monthly equity
returns. The implicit copula here is a ``regression copula'' process with respect to the covariates as in~\cite{KleSmi2019}. 
The copula model forms
a distributional regression~\citep{KleKneLan2015,kneib2021}, where the five factors affect the entire distribution
of equity returns, not just its first or other moments. 
In both applications the implicit copulas are of dimension equal to the number of 
observations, so that they are high-dimensional. Nevertheless, their auxiliary representation allows for likelihood-based estimation of the copula parameters.
In both examples the marginal distribution of the response variables exhibit strong asymmetries.

The overview is organized as follows. Section~\ref{sec:implicitcops} introduces
general
copula models, and then implicit copulas specifically. Their interpretation as transformations and specifications for variables that are continuous, discrete or mixed 
are also discussed. Section~\ref{sec:eandsecop} covers elliptical and skew-elliptical copulas,
including the Gaussian, $t$, skew $t$ and factor copulas.
Implicit copulas that capture serial dependence in time series data are covered in Section~\ref{sec:timeseries}. Section~\ref{sec:mtimeseries} extends these to implicit copulas
that capture both serial and cross-sectional dependence in multivariate time series. 
Section~\ref{sec:regcopprocess} covers regression copula processes, with the 
implicit copula constructed
from a
regularized linear regression given in detail.
It is shown that when this copula is combined with flexible marginals, it defines a promising new distributional regression model.
Last, Section~\ref{sec:discuss} discusses the advantages of using 
implicit copula models for modeling data, and future directions.

%% file: sec2.tex
\section{Implicit copulas}\label{sec:implicitcops}

\subsection{Copula models in general}
All copula models are based on the theorem of~\cite{sklar59} (i.e. ``Sklar's theorem''), which states that 
for every random vector $\bm{Y}=(Y_1,\ldots,Y_m)^\top$ with distribution function $F_Y$ and
marginals $F_{Y_1},\ldots,F_{Y_m}$, there exists a ``copula function'' $C:[0,1]^m \rightarrow [0,1]$, such that
\begin{equation}
	F_Y(\bm{y})=C(F_{Y_1}(y_1),\ldots,F_{Y_m}(y_m))\,,\label{eq:sklarcdf}
\end{equation}
where $\bm{y}=(y_1,\ldots,y_m)^\top$. The copula function $C$ is  a well-defined distribution
function for a random vector $\bm{U}=(U_1,\ldots,U_m)^\top$ on the unit cube with uniform marginal distributions.
To  construct a copula model, select   
$F_{Y_1},\ldots,F_{Y_m}$ (i.e. the ``marginal models'') and a copula function 
$C$, to define $F_Y$ via~\eqref{eq:sklarcdf}. 

\subsubsection{Continuous case}
If all the elements of $\bm{Y}$ are 
continuous, then differentiating through~\eqref{eq:sklarcdf} gives the density 
\begin{equation}
	f_Y(\bm{y})=\frac{\partial^m}{\partial y_1\cdots \partial y_m}F_Y(\yvec)=
	c(F_{Y_1}(y_1),\ldots,F_{Y_m}(y_m))\prod_{j=1}^m f_{Y_j}(y_j)\,,\label{eq:sklarpdf}
\end{equation}
where $f_{Y_j}=\frac{\partial}{\partial y_j} F_{Y_j}$, and
$c(\bm{u})=\frac{\partial^m}{\partial u_1\cdots \partial u_m} C(\bm{u})$ is widely called the 
``copula density''
	 with $\bm{u}=(u_1,\ldots,u_m)^\top$. (Throughout this paper the notation $c(\uvec)$ and $c(u_1,u_2,\ldots,u_m)$
	 	are used interchangeably, as are $C(\uvec)$ and $C(u_1,u_2,\ldots,u_m)$.) The decomposition at~\eqref{eq:sklarpdf} is used to specify the likelihood of a continuous 
response vector $\bm{Y}$ in a statistical model. 

\subsubsection{Discrete case}\label{sssec:discretecase}
If all the elements of $\bm{Y}$ are discrete-valued (e.g. as with ordinal or binary data)
the probability mass function is obtained by differencing over the elements of $\bm{Y}$
as follows. Let $b_j=F_{Y_j}(y_j)$ and $a_j=F_{Y_j}(y_j^-)$ be the left-hand limit 
of $F_j$ at $y_j$ (which is $a_j=F_{Y_j}(y_j-1)$ for ordinal $Y_j$). Then the 
mass function is
\begin{equation}
	f_Y(\bm{y})=\mbox{Pr}(Y_1=y_1,\ldots,Y_m=y_m)=\Delta_{a_1}^{b_1}\Delta_{a_2}^{b_2}\cdots
	\Delta_{a_m}^{b_m}C(\vvec)\,,\label{eq:copmass}
\end{equation}
where $\vvec=(v_1,\ldots,v_m)^\top$ is a differencing vector, and the notation 
\begin{eqnarray*}
\lefteqn{\Delta_{a_j}^{b_j} C(u_1,\ldots,u_{j-1},v_j,u_{j+1},\ldots,u_m)}&&\\
&= &C(u_1,\ldots,u_{j-1},b_j,u_{j+1},\ldots,u_m)-C(u_1,\ldots,u_{j-1},a_j,u_{j+1},\ldots,u_m)\,.
\end{eqnarray*}
Evaluating the mass function at~\eqref{eq:copmass} is an $O(2^m)$ computation, so that its direct evaluation 
is impractical for high
values of $m$ when undertaking likelihood-based estimation~\citep{nikoloulopoulos2013}.
 One solution suggested by~\cite{smithkhaled2012} is to consider the joint 
distribution of $(\bm{Y},\bm{U})$. To do so, note that when $Y_j$ is discrete,
$F_{Y_j}$ is a many-to-one function and $Y_j|U_j$ is a degenerate distribution with density
 $f(y_j|u_j)=\mathds{1}(a_j \leq u_j < b_j)$, where the indicator function $\mathds{1}(X)=1$ if $X$ is true, and zero 
 otherwise. (An alternative notation is to use the Dirac delta function, with $f(y_j|u_j)=\delta_{y_j}(F_{Y_j}^-(u_j))$ where
 	$F_{Y_j}^-$ is the quantile function of $Y_j$.) Then the mixed density of $(\bm{Y},\bm{U})$ is
\begin{equation}
	f_{Y,U}(\yvec,\uvec)=f(\yvec|\uvec)c(\uvec)=\prod_{j=1}^m \left\{ \mathds{1}(a_j \leq u_j < b_j)\right\}c(\uvec)\,.\label{eq:extdensity}
\end{equation} 
Marginalizing over $\bm{U}$ gives the probability
mass function at~\eqref{eq:copmass} (i.e. $f_Y(\yvec)=\int f_{Y,U}(\yvec,\uvec)\mbox{d}\uvec$); see Proposition~1 in~\cite{smithkhaled2012}.

Equation~\eqref{eq:extdensity} can be used to define an ``extended likelihood'' for 
estimation using computational methods for latent variables, where
the observations on $\bm{U}$ are the latents. This has two
advantages. First, the $O(2^m)$ computation at~\eqref{eq:copmass} is avoided, allowing
estimation for higher values of $m$. Second,
only the copula density $c$ is required and not the copula function $C$, which is
an advantage for some copulas where only $c$ can be computed,
as is the case with most vine copulas~\citep{joe1996,AasCzaFriBak2009}.
Bayesian data augmentation can be used based on~\eqref{eq:extdensity}, and 
evaluated using Markov chain Monte Carlo (MCMC)
 as in~\cite{smithkhaled2012} or variational Bayes methods as in~\cite{loaiza2019VB}.
 The latter is particularly attractive, because it allows for the estimation of 
discrete-margined copulas of very high dimensions, with examples up to $m=792$ presented
by these authors.

\subsubsection{Mixed cases}
If some elements of $\bm{Y}$ are continuous and others discrete, then 
 $f_Y$ is often called a ``mixed density''. In this case, an extended likelihood can be 
constructed from the distribution of $\bm{Y}$ joint with the elements of $\bm{U}$ 
that correspond only to the discrete variables; see~\citet[Sec.6]{smithkhaled2012}. 
Similarly,
if some individual elements $Y_j$ have distributions that are mixtures
of continuous and discrete distributions (such as a zero-inflated 
continuous distribution) then an extended likelihood can also be constructed
for this case; see~\cite{gunawan2020} for how to do so.

\subsection{The basic idea of an implicit copula}\label{sec:basicidea}
\citet[p.190]{McNFreEmb2005} use the term ``implicit copula'' for the
copula that is implicit in the multivariate distribution of a continuous 
 random vector $\bm{Z}=(Z_1,\ldots,Z_m)^\top$. It is obtained by inverting Sklar's theorem, which~\citet[p.51]{nelsen06} calls the ``inversion
 method'', so that copulas derived in this fashion are also called ``inversion copulas'' (e.g.~\cite{smithman2018}). 
 If $\bm{Z}$ has distribution function $F_Z$ with marginals $F_{Z_1},\ldots,
 F_{Z_m}$, then its implicit copula function is
 \begin{equation}
 	C_Z(\uvec)=F_Z\left(F_{Z_1}^{-1}(u_1),\ldots,F_{Z_m}^{-1}(u_m)\right)\,.\label{eq:icopcdf}
 \end{equation}
Differentiating with respect to $\uvec$ gives the implicit copula density
\begin{equation}
	c_Z(\uvec)=\frac{\partial^m}{\partial u_1\cdots \partial u_m}C(\uvec)=\frac{f_Z(\zvec)}{\prod_{j=1}^m f_{Z_j}(z_j)}\,,\label{eq:icoppdf}
\end{equation}
where $\zvec=(z_1,\ldots,z_m)^\top$ is a function of $\uvec$ with elements $z_j=F_{Z_j}^{-1}(u_j)$ for 
$j=1,\ldots,m$. The implicit copula function $C_Z$ and density $c_Z$ above can be 
employed in~\eqref{eq:sklarcdf}, \eqref{eq:sklarpdf} and~\eqref{eq:copmass}. Thus,
an implicit copula model uses Sklar's theorem twice: once to form the joint
distribution $F_Y$ with arbitrary marginals, and a second time to construct the implicit copula from the joint distribution $F_Z$.

Because implicit copulas are an immediate consequence of Sklar's theorem, they have
a long history.  Early uses for modelling data include~\cite{ruschendorf1976} and~\cite{deheuvels1979}, who both construct a non-parametric
implicit copula from the empirical distribution function (although neither called it
a copula). \cite{ruschendorf2009} gives
an overview of the early developments of implicit copulas, pointing out that many transformation-based multivariate models---which themselves have a long history---are 
also copula models based on implicit copulas (although in the early literature this was often unrecognized and
the term ``copula'' not used).

Note that only
a continuous distribution $F_Z$ is used to construct an implicit copula here. 
This is because the implicit copula of a discrete distribution $F_Z$ is not 
unique~\citep{genest2007}.

\subsection{Implicit copulas as transformations}
One way to look at all copula models is that they are a transformation 
from $\bm{Y}$ to $\bm{U}=(U_1,\ldots,U_m)^\top \in[0,1]^m$.  The key observation is that
it is usually easier to capture multivariate dependence using $C$ on the vector space $[0,1]^m$,
rather than directly on the domain of the original vector $\bm{Y}$.
Implicit copulas go one step further, with a second transformation from $\bm{U}$ to 
$\bm{Z}=(F_{Z_1}^{-1}(U_1),\ldots,F_{Z_m}^{-1}(U_m))^\top$, and then capture the dependence 
structure using the distribution $F_Z$. Table~\ref{tab:icop} provides
a summary of these transformations, 
along with the marginal and joint distribution and density/mass functions of
$\bm{Y}$.
Throughout this paper,
the vector $\bm{U}$ is referred to as the ``copula vector'' and $\bm{Z}$ as the ``auxiliary vector'' (the latter is also called a ``pseudo vector'' in \cite{smith+k19}). Simulation from an implicit copula model is straightforward if 
$F_Z$ is tractable using Algorithm~\ref{alg:simicop}, which produces a draw $\yvec\sim F_Y$.

\begin{algorithm}
	\caption{\em (Random iterate generation from an implicit copula model)}
	\label{alg:simicop}
	\begin{itemize}
\item[1.] Generate $\zvec=(z_1,\ldots,z_m)^\top \sim F_Z$
\item[2.] For $j=1,\ldots,m$, set $u_j=F_{Z_j}(z_j)$, and $\uvec=(u_1,\ldots,u_m)^\top$
\item[3.] For $j=1,\ldots,m$, set $y_j=F_{Y_j}^{-1}(u_j)$, and $\yvec=(y_1,\ldots,y_m)^\top$
	\end{itemize}
\end{algorithm}

Notice that 
the transformation $U_j=F_{Z_j}(Z_j)\sim \mbox{Uniform}[0,1]$ removes all features
of the marginal distribution of $Z_j$.  This becomes an important observation for
establishing parameter identification when constructing 
implicit copulas, as discussed in Sections~\ref{sec:timeseries},~\ref{sec:mtimeseries} and~\ref{sec:regcopprocess}. 
\begin{sidewaystable}[p]
	\caption{Transformational relationships between observational vector $\bm{Y}$, copula vector $\bm{U}$ 
		and auxiliary vector $\bm{Z}$ for an implicit copula}\label{tab:icop}
\begingroup
\renewcommand{\arraystretch}{1.25}
\begin{center}	
		\begin{tabular}{llll}\hline \hline
			& Observational  &Copula  &Auxiliary \\ \cline{2-4}
			\multirow{2}{*}{Random Variable} &Continuous $Y_j$ &$U_j=F_{Y_j}(Y_j)$ &$Z_j=F_{Z_j}^{-1}(U_j)$ \\
			 &Discrete $Y_j$ &$F_{Y_j}(Y_j^-)\leq U_j < F_{Y_j}(Y_j)$ 
			&$F_{Z_j}^{-1}(F_{Y_j}(Y_j^-))\leq Z_j < F_{Z_j}^{-1}(F_{Y_j}(Y_j))$ \\
			Domain &${\cal D}_{Y_1} \times \cdots \times {\cal D}_{Y_m}$ &
			$[0,1]^m$ & ${\cal D}_{Z_1} \times \cdots \times {\cal D}_{Z_m}$ \\
			Marginal Distribution &$F_{Y_j}$ &Uniform &$F_{Z_j}$\\
			Joint Distribution &$F_Y(\yvec)=C(\uvec)$ &$C(\uvec)=F_Z(F_{Z_1}^{-1}(u_1),\ldots,F_{Z_m}^{-1}(u_m))$ 
			&$F_Z$ \\
			Joint Density/Mass &&& \\
			\mbox{  } ($Y_j$ Continuous)  &$f_Y(\bm{y})=c(\uvec)\prod_{j=1}^m f_{Y_j}(y_j)$  
			&\multirow{2}{*}{$c(\bm{u})=\frac{f_Z(\zvec)}{\prod_{j=1}^m f_{Z_j}(z_j)}$} &\multirow{2}{*}{$f_Z(\zvec)$}\\
			\mbox{  } ($Y_j$ Discrete) &$f_Y(\bm{y})=\Delta_{a_1}^{b_1}\Delta_{a_2}^{b_2}\cdots\Delta_{a_m}^{b_m}C(\vvec)$&&\\ \hline \hline
		\end{tabular}
	\end{center}
\endgroup
				
	\noindent The joint distribution and density/mass functions 
	of $\bm{Y}=(Y_1,\ldots,Y_m)^\top$, $\bm{U}=(U_1,\ldots,U_m)^\top$ and $\bm{Z}=(Z_1,\ldots,Z_m)^\top$ are given. The joint
	density of $\bm{Y}$ is given separately when all the elements are continuous and when all the elements are discrete. 
	When some elements are discrete and others continuous,
	the mixed density is given in~\citet[Sec.6]{smithkhaled2012}. In this table, ${\cal D}_{Y_j}$ is the domain of $Y_j$, and ${\cal D}_{Z_j}$ is the domain of $Z_j$.
\end{sidewaystable}

\subsection{An alternative extended likelihood}\label{ssec:extlike}
For the case where the elements of $\bm{Y}$ are discrete-valued, for an implicit copula 
model there exists an alternative extended likelihood based on the joint density of $(\bm{Y},\bm{Z})$,
rather than that of $(\bm{Y},\bm{U})$ given previously at~\eqref{eq:extdensity}. This alternative 
joint density is
\begin{equation}
	f_{Y,Z}(\yvec,\zvec)=f(\yvec|\zvec)f_Z(\zvec)=\prod_{j=1}^m \left\{\mathds{1}\left(F_{Z_j}^{-1}(a_j) \leq z_j < F_{Z_j}^{-1}(b_j)\right)\right\} f_Z(\zvec)\,,\label{eq:extdensityz}
\end{equation}
with $a_j,b_j$ as defined above in Section~\ref{sssec:discretecase}. Marginalizing over $\bm{Z}$ produces the 
probability mass function at~\eqref{eq:copmass}; i.e. $f_Y(\yvec)=\int f_{Y,Z}(\yvec,\zvec)\mbox{d}\zvec$. An advantage is that it is often simpler
to use computational methods to estimate an implicit copula
using~\eqref{eq:extdensityz} rather than~\eqref{eq:extdensity}. 
Moreover, an extended likelihood is also easily defined for
vectors $\bm{Y}$ with combinations of discrete, continuous or even mixed valued elements, by simplifying~\eqref{eq:extdensityz} 
to only include elements of $\bm{Z}$ that correspond to the non-continuous valued variables.

Bayesian data augmentation is a suitable method for estimation using
this extended likelihood. Here, values for
$\zvec$ are generated in an MCMC sampling scheme to evaluate an ``augmented posterior'' 
proportional to the extended likelihood
multiplied by a parameter prior. This has been used to estimate the elliptical 
and skew elliptical copulas discussed in Section~\ref{sec:eandsecop} below. For example,
\cite{pitt2006} do so
for a Gaussian copula, while~\cite{danaher2011} do so for the $t$ copula, and~\cite{smith2012} for the skew $t$ copula. \cite{hoff2007} considered
the extended likelihood above using empirical marginals and rank data, \cite{danaher2011} and~\cite{dobra2011}
provide early applications to higher dimensional Gaussian $F_Z$. Last, the multivariate probit model is a
Gaussian copula model, and the popular approach of~\cite{chib1998} is a special case of these data augmentation algorithms.

%% file: sec3.tex
\section{Elliptical and Skew Elliptical Copulas}\label{sec:eandsecop}
In practice, parametric copulas $C(\uvec;\thetavec)$ with parameter vector $\thetavec$ are almost always used
in statistical modelling, with~\cite{McNFreEmb2005}, \cite{nelsen06} and~\cite{joe2014dependence}
giving overviews of choices.
However, the implicit copulas of elliptical distributions, and more recently skew elliptical distributions, 
 are common choices for capturing dependence in many applications. An
 attractive feature is that because elliptical and skew elliptical
 distributions are closed under marginalization, so are their implicit copulas. 

\subsection{Elliptical copulas}\label{sec:elliptical}
\subsubsection{Gaussian copula}\label{sssec:gcop}
The simplest and most popular elliptical copula is the ``Gaussian copula'', 
which is constructed from
$\bm{Z}\sim N_m(\bm{0},\Omega)$ with $\Omega$  an $m\times m$ correlation matrix. If $\Phi_m(\cdot;\avec,\Omega)$ denotes an $N_m(\avec,\Omega)$ distribution function, 
and $\Phi(\cdot)$ a $N(0,1)$ distribution function, then from~\eqref{eq:icopcdf} the Gaussian copula
function is 
\begin{equation*}
	C_{\tiny{\mbox Ga}}(\uvec;\Omega)=\Phi_m\left(\Phi^{-1}(u_1),\ldots,\Phi^{-1}(u_m);\bm{0},\Omega\right)\,.
\end{equation*}
If $\phi_m(\cdot;\avec,\Omega)$ is a $N_m(\avec,\Omega)$ density, and $\phi$ is a standard normal density, then
plugging the Gaussian densities into~\eqref{eq:icoppdf} gives the Gaussian copula density
\begin{equation*}
	c_{\tiny{\mbox Ga}}(\uvec;\Omega) = \phi_m(\zvec;\bm{0},\Omega)/\prod_{j=1}^m \phi(z_j)\\
	= |\Omega|^{-1/2}\exp\left\{-\frac{1}{2}\zvec^\top (\Omega^{-1}-I_m)\zvec\right\}\,,
\end{equation*}
with $\zvec=(\Phi^{-1}(u_1),\ldots,\Phi^{-1}(u_m))^\top$. 

There are a number of immediate observations on the Gaussian copula.
First, the auxiliary vector $\bm{Z}$ has a distribution with a zero mean and unit marginal variances. This is because information about the first two marginal moments of
$Z_j$ are lost in the transformation $U_j=F_{Z_j}(Z_j)$ and
are unidentified
in the copula density. Second, adopting any constant mean value (other than zero) and marginal
variances (other than unit values) for $\bm{Z}$ produces the same 
Gaussian copula $C_{\tiny{\mbox Ga}}$. Third, closure under marginalization
means
that if $\bm{U}$ has
distribution function $C_{\tiny{\mbox Ga}}(\uvec;\Omega)$, then any subset $\bm{U}_0$ of elements of $\bm{U}$ has distribution function $C_{\tiny{\mbox Ga}}(\uvec^0;\Omega^0)$,
where $\Omega^0$ is a correlation matrix made up of the corresponding rows and 
columns of $\Omega$. 

A fourth observation is that any parametric correlation
structure for $\bm{Z}$ is inherited by the Gaussian copula. It is 
this property that has led the widespread adoption of Gaussian copula models for
modeling
time series~\citep{cario1996}, longitudinal~\citep{lambert2002}, cross-sectional~\citep{murray2013} and   
spatial~\citep{bai2014,hughes2015} data. The Gaussian copula has a long history, particularly when
formed implicitly via transformation (e.g.~\cite{LiHammond}), although 
some early and influential mentions include~\cite{joe1993}, \cite{clemen1999} and~\cite{wang1999}, while~\cite{li2000}
popularized its use in finance. 
A comprehensive overview of the Gaussian copula and its properties is given
by~\cite{Song2000}.

\subsubsection{Other elliptical copulas}
\cite{fang2002} and~\cite{embrechts2002} use an elliptical distribution for $\bm{Z}$,
and study the resulting class of ``elliptical copulas''. When combined
with choices for the marginals of $\bm{Y}$ in a copula model, \cite{fang2002} call the distribution $F_Y$ ``meta-elliptical'',
and an overview of their dependence properties is given by~\cite{abdous2005}. After the Gaussian copula, the
most popular elliptical copula is the $t$ copula, where a multivariate $t$ distribution with 
degrees of freedom $\nu>0$
is adopted for $\bm{Z}$. 
\cite{embrechts2002} and~\cite{venter2003} study this copula,
and the main advantage is that
it can capture higher dependence in extreme values, which is important 
for financial and actuarial variables.
A lesser known property is that values of $\nu$ close to zero 
allow for positive dependence between squared 
elements of $\bm{Y}$. This is useful for capturing the serial 
dependence in heteroscedastic time series, such as equity returns in finance; 
see~\cite{loaiza2018hetero} and~\cite{bladt2021}.

\subsection{Skew elliptical copulas}\label{sec:skewtcop}
\subsubsection{Overview}
Elliptical copulas exhibit radial symmetry, where the distributions of 
$(U_i,U_j)$ and $(1-U_i,1-U_j)$ are the same. 
Yet there are applications where this is unrealistic,
including for the dependence between equity returns~\citep{longin2001,ang2002asymmetric} and regional
electricity spot prices~\citep{smith2012}. The implicit copulas of
skew elliptical distributions~\citep{genton2004} allow for 
asymmetric pairwise dependence, with the most common
being those constructed from the differing skew $t$ distributions.
\cite{demarta2005} were the first to construct an implicit copula from a skew $t$ distribution (i.e. a ``skew $t$ copula''),
for which they used a special case of the generalized hyperbolic distribution,
and~\cite{chan2010} do so for an adjustment of the skew normal distribution
of~\cite{azzalini1996}.
The most popular variants of the skew $t$ distribution are those of~\cite{Azz2003} and~\cite{sahu2003},
which
share a similar conditionally Gaussian representation.
\cite{smith2012} show how to construct
implicit copulas from these latter two
skew $t$ distributions, and estimate them using MCMC. 
\cite{yoshiba2018}
considers maximum likelihood estimation for the skew $t$ copula constructed
from the distribution of~\cite{Azz2003}, and~\cite{oh2020dynamic} consider a dynamic extension of the
skew $t$ copula of~\cite{demarta2005} for high dimensions.

\subsubsection{Skew $t$ copula}
Write $t_d(\avec,\Omega,\nu)$ for a $d$-dimensional $t$ distribution with location 
$\avec$, scale matrix $\Omega$ and degrees of freedom $\nu$, with density $f_t(\cdot;\avec,\Omega,\nu)$. 
Let $\bm{X}$ and $\bm{Q}$ be $(m\times 1)$ vectors with joint distribution
\begin{equation}
	\left(\begin{array}{c} \bm{X}\\ \bm{Q} \end{array}\right) \sim t_{2m}\left(
	\left(\begin{array}{c} \bm{0}\\ \bm{0} \end{array}\right),
	\Omega=\left(\begin{array}{cc} \Gamma+D^2 &D\\ D&I \end{array}\right),
	\nu
	\right)\,.\label{eq:skewtjoint}
\end{equation}
Here, $D=\mbox{diag}(\delta_1,\ldots,\delta_m)$ is a diagonal matrix and $\Gamma$ is positive
definite. Then the skew $t$ distribution of~\cite{sahu2003} (with location parameter equal to zero) is given by $\bm{Z}=(\bm{X}|\bm{Q}>\bm{0})$, which has density 
\begin{equation}\label{eq:skewtpdf}
	f_{\mbox{\tiny St}}(\zvec;\Gamma,D,\nu)=
	\frac{2^m}{|\Gamma+D^2|^{1/2}}f_t\left((\Gamma+D^2)^{-1/2}\zvec ;\bm{0},I_m,\nu\right)\mbox{Pr}(\bm{V}>\bm{0};\zvec)
\end{equation} 
where $\bm{V}\sim t_m\left(D(\Gamma+D^2)^{-1}\zvec,\frac{S(\zvec)+\nu}{m+\nu}(I-D(\Gamma+D^2)^{-1}D,m+\nu\right)$
and $S(\zvec)=\zvec'(\Gamma+D^2)^{-1}\zvec$. This manner of constructing a skew $t$ distribution is called ``hidden conditioning''
because $\bm{Q}$ is latent. The skew $t$ distribution of~\cite{Azz2003} is constructed
	in a similar way, but where $\bm{Q}$ is a scalar.

The parameter $\deltavec=(\delta_1,\ldots,\delta_m)^\top$ controls the level of asymmetry in
the distribution of $\bm{Z}$,
but in the implicit copula it controls the level of asymmetric dependence. This is a key observation as to 
why skew $t$ copulas have strong potential for applied modeling.  
To construct this copula, first fix the leading diagonal elements of $\Gamma$
to ones (i.e. restrict $\Gamma$ to be a correlation matrix), and
note that the marginal of $Z_j$ is also a skew $t$ distribution with density
$f_{\mbox{\tiny St}}(z_j;1,\delta_j,\nu)$. Then, the copula function and density 
are given by~\eqref{eq:icopcdf} and~\eqref{eq:icoppdf}, respectively. 
These require computation of the distribution 
function $F_{Z_j}(z_j)=\int_{-\infty}^{z_j}f_{\mbox{\tiny St}}(z_j';1,\delta_j,\nu)\mbox{d}z_j'$ and its inverse (i.e. the quantile function)
which can either be undertaken numerically using standard methods, or 
using the interpolation approach outlined in~\ref{app:A} for large datasets.
Simulation from a skew $t$ copula model is straightforward using~\eqref{eq:skewtjoint}
and the representation of a $t$ distribution as Gaussian conditional on a Gamma variate. To do so,
at Step~1 of Algorithm~1 generate a draw $\zvec\sim F_Z$ by drawing
sequentially as follows:
\begin{itemize}
	\item[] Step~1(a) Generate $w\sim \mbox{Gamma}(\nu/2,\nu/2)$,
	\item[] Step~1(b) Generate $\bm{q}\sim N_m(\bm{0},\frac{1}{w}I_m)$ constrained to $\bm{Q}>\bm{0}$,
	\item[] Step~1(c) Generate $\bm{z}\sim N_m(D\bm{q},\frac{1}{w}\Gamma)$.
\end{itemize} 

A computational bottleneck for the evaluation of the skew $t$ copula
density is the evaluation of 
the multivariate integral $\mbox{Pr}(\bm{V}>\bm{0};\zvec)$ at~\eqref{eq:skewtpdf}.

However, this
can be avoided in likelihood-based estimation by considering the tractable conditionally
Gaussian representation motivated by~\eqref{eq:skewtjoint}. Let $W\sim \mbox{Gamma}(\nu/2,
\nu/2)$, then consider the joint distribution of $(\bm{X},\bm{Q},W|\bm{Q}>\bm{0})$ with 
density 
\begin{equation}
f(\xvec,\qvec,w|\bm{q}>\bm{0})\propto f(\xvec|\qvec,w)f(\qvec|w)\mathds{1}(\qvec>\bm{0})f(w)\,,
\label{eq:skewtext}
\end{equation}
where $(\bm{X}|\bm{Q}=\qvec,W=w)\sim N_m(D\qvec,\frac{1}{w}\Gamma)$ and
$(\bm{Q}|W=w)\sim N_m(\bm{0},\frac{1}{w}I_m)$.
Marginalizing out 
$(\qvec,w)$ gives the skew $t$ density at~\eqref{eq:skewtpdf} in $\xvec$. \cite{smith2012} use this
feature to design Bayesian data augmentation algorithms for the skew $t$ copula
that generate $(\qvec,w)$ as latent variables in Markov chain Monte Carlo (MCMC) sampling schemes
for both continuous-valued and discrete-valued $\bm{Y}$. 

The density of the~\cite{Azz2003} skew $t$ distribution does not feature the multivariate
probability term $\mbox{Pr}(\bm{V}>0)$, so that it is easier to evaluate its implicit copula density, as in~\cite{yoshiba2018}. 
But when computing the Bayesian posterior using data augmentation
it makes little difference, because the copula density is never evaluated directly. 

\subsection{Factor copulas}
To capture dependence in high dimensions, ``factor copulas'' are increasingly popular, and 
there are two main types in the literature.
The first links a small number of independent factors
by a pair-copula construction to produce a higher dimensional copula, as  proposed by~\cite{krupskii2013}. 
Flexibility is obtained by
using different bivariate copulas for the pair-copulas and a different number of factors, with
applications and extensions
found in~\cite{nikoloulopoulos2015factor,mazo2016,schamberger2017,tan2019} and~\cite{krupskii2020}.
In general, this type of factor copula is not an implicit copula.
The second type of factor copula is the  implicit copula of a traditional
elliptical or skew-elliptical factor model. 
This type of copula emerged in the finance literature for low-dimensional 
applications~\citep{laurent2005}, but is
increasingly used to model dynamic dependence in high 
dimensions; see~\cite{creal2015,oh2017,oh2018} and~\cite{oh2020dynamic}. Estimation
issues grow with the dimension and complexity of the copula, and this remains an active
field of research. 

\subsubsection{Gaussian static factor copula}
One of the simplest factor copulas is a Gaussian static factor copula, which \cite{laurent2005} suggest
for a single factor, and~\cite{murray2013} consider for a larger number of factors. The multiple factor
copula can be defined as follows.
Let $\widetilde{\bm{Z}} \sim N_m(\bm{0},\Lambda \Lambda^\top +D)$, where $\Lambda=\{\lambda_{j,k}\}$ is an $m\times p$ matrix of factor loadings, $D=\mbox{diag}(d_1,\ldots,d_m)$ is a diagonal matrix of idiosyncratic 
variations,
and typically $p<<m$. 
The implicit copula of $\widetilde{\bm{Z}}$ is a Gaussian copula, as outlined
in Section~\ref{sssec:gcop}. To derive the parameter matrix $\Omega$, set the diagonal matrix 
\[
S=\mbox{diag}(\Lambda \Lambda^\top +D)=\mbox{diag}\left(\sum_{k=1}^p \lambda_{1,k}^2+d_1,\ldots,\sum_{k=1}^p \lambda_{m,k}^2+d_m\right)\,,
\]
then $\bm{Z}=S^{-1/2}\widetilde{\bm{Z}}$, so that $\Omega=S^{-1/2}(\Lambda \Lambda^\top + D)S^{-1/2}$.

\cite{murray2013} identify the loadings and idiosyncratic variations by 
setting $D=I$, the upper triangular elements of $\Lambda$ to zero
and the leading diagonal elements to positive values $\lambda_{i,i}>0$.
The copula parameters are then 
$\thetavec=(\mbox{vecl}(\Lambda),d_1,\ldots,d_m)$, where 
$\mbox{vecl}(\Lambda)$ is the half-vectorization
operator applied to the lower triangle 
of the rectangular matrix $\Lambda$. 
In the non-copula factor model literature, there are alternative
ways to identify $\Lambda$ and $D$ \citep{kaufmann2017,fruhwirth2018}, and 
similar restrictions may be adapted for the correlation matrix $\Omega$ as well.
In a Bayesian framework, priors also have to be adopted for $\Lambda$ and $D$,
and these can be used to provide further regularization as in~\cite{murray2013}
and elsewhere.

Simulation from this factor copula model is fast using the latent variable representation of the factor structure
given by $\etavec\sim N_p(\bm{0},I)$ and $\widetilde{\bm{Z}}|\etavec \sim N_m(\Lambda \etavec,D)$. To do so, at Step~1 of 
Algorithm~\ref{alg:simicop} generate a draw $\zvec\sim F_Z$ by drawing sequentially as follows:
\begin{itemize}
	\item[] Step 1(a) Generate $\etavec \sim N_p(\bm{0},I)$ and $\epsilonvec\sim N_m(\bm{0},D)$,
	\item[] Step 1(b) Set $\widetilde{\bm{z}}=\Lambda \etavec + \epsilonvec$,
	\item[] Step 1(c) Set $\zvec = S^{-1/2} \widetilde{\bm{z}}$.
\end{itemize}
 

%% file: sec4.tex
\section{Time series}\label{sec:timeseries}
Copulas have been used extensively to capture the cross-sectional dependence in multivariate time series; see~\cite{patton2012} for a review. However, they can also be used
to capture the serial dependence in a univariate series. The resulting time series
models are 
extremely flexible, and there are many potential applications to continuous, discrete or
mixed data.

\subsection{Time series copula models}
If $\bm{Y}=(Y_1,\ldots,Y_T)^\top$ is a time series 
vector, 
then the copula $C$ at~\eqref{eq:sklarcdf} with $m=T$ captures
the serial dependence in the series and is called a ``time series copula''.
While there has been less work on time series copulas than those used to capture
cross-sectional dependence, they are increasingly being used for both time series data
(where there is a single observation on the vector $\bm{Y}$) and 
longitudinal data (where there are multiple observations on the vector $\bm{Y}$).
Early contributions
 include~\cite{darsow1992}, \citet[Ch.8]{joe97}, \cite{freeswang2005}, 
\cite{chen2006tscop}, \cite{ibragimov2009} and~\cite{beare2010} for Markov processes,
\cite{Wilson2010} for the implicit copulas of Gaussian processes popular in machine learning,
and \cite{smith2010vine} for
vine copulas that exploit the time ordering of the
elements of $\bm{Y}$.

\subsubsection{Decomposition}
For a continuous-valued stochastic process $\{Y_t\}$, denote the copula model for the joint density of time series variables
$\bm{Y}_{1:t}=(Y_1,\ldots,Y_t)^\top$ as
\[ 
f_{Y_{1:t}}(y_1,\ldots,y_t)=c_{1:t}(u_1,\ldots,u_t)\prod_{s=1}^t f_{Y_s}(y_s)\,,
\]
where $c_{1:t}$ is a $t$-dimensional copula density that defines a
{\em copula process} for stochastic process $\{U_t\}$, with $U_t=F_{Y_t}(Y_t)$. Then
the conditional distribution $Y_{t+1}|\bm{Y}_{1:t}$ 
has density
\begin{eqnarray}
	f_{Y_{t+1|1:t}}(y_{t+1}|y_{1},\ldots,y_t) &=&
	\frac{f_{Y_{1:t+1}}(y_1,\ldots,y_{t+1})}{f_{Y_{1:t}}(y_1,\ldots,y_{t})}=\frac{c_{1:t+1}(u_1,\ldots,u_{t+1})}{c_{1:t}(u_1,\ldots,u_{t})}f_{Y_{t+1}}(y_{t+1})\nonumber \\
	&= &f_{U_{t+1|1:t}}(u_{t+1}|u_1,\ldots,u_t)f_{Y_{t+1}}(y_{t+1})\,.\label{eq:tscopcondpdf}
\end{eqnarray}
Here, $f_{U_{t+1|1:t}}$ is the density of $(U_{t+1}|U_1,\ldots,U_t)$, which is not uniform on $[0,1]$ (whereas the marginal distribution of $U_{t+1}$ is uniform on $[0,1]$).
This conditional density can be used to form predictions from the copula model. It can also be used in likelihood-based estimation because
$f_Y(\yvec)=
\prod_{t=2}^T\left\{ f_{U_{t|1:t-1}}(u_{t}|u_1,\ldots,u_{t-1})f_{Y_{t}}(y_{t})\right\}f_{Y_1}(y_1)$, with
$\yvec=(y_1,\ldots,y_T)^\top$,
which can be computed
efficiently for many choices of copula $c_{1:T}$.
In drawable vine 
copulas (D-vines) $f_{U_{t+1|1:t}}$ is further decomposed into a product of bivariate
copulas called ``pair-copulas''~\citep{AasCzaFriBak2009}, allowing for a flexible
representation of the serial dependence structure, as discussed by~\cite{smith2010vine},
\cite{beare2015}, \cite{smith2015}, \cite{loaiza2018hetero}, \cite{bladt2021} and others.

\subsubsection{Selection of marginal distributions}
For longitudinal data with a sufficient number of
observations on $\bm{Y}$, it is possible to estimate the marginal distribution
functions $F_{Y_1},\ldots,
F_{Y_T}$ at~\eqref{eq:icopcdf} separately as in~\cite{smith2010vine}. 
But for time series data it is necessary to impose some structure on these marginal densities. For example, \cite{freeswang2005,freeswang2006}
employ generalized linear
regression models with time-based covariates in an actuarial setting. 
In the absence of common covariates, the marginals may be assumed time-invariant, so that $F_{Y_t}\equiv G$ for all $t$ as in~\cite{chen2006tscop} and~\cite{smith2015}. 
Flexible marginals, such as a skew $t$
distribution, or non-parametric estimators
such as smoothed empirical distribution functions or kernel density estimators, can be used.

\subsubsection{Discrete time series data}
Time series copulas can also be used for discrete-valued data; see~\citet[Ch.8]{joe97} for
an early exploration of such models. \citet[Sec.5]{smithkhaled2012} do 
so for longitudinal data using the extended likelihood at~\eqref{eq:extdensity} and the 
copula decomposition above, so that for $\yvec=(y_1,\ldots,y_T)^\top$ and $\uvec=(u_1,\ldots,u_T)^\top$,
\begin{eqnarray}
	f_{Y,U}(\yvec,\uvec) &= &\prod_{t=1}^T\left\{\mathds{1}(a_t \leq u_t < b_t)\right\}c_{1:T}(\uvec) \nonumber\\
	&= &\prod_{t=2}^T\left\{ f_{U_{t|1:t-1}}(u_{t}|u_1,\ldots,u_{t-1}) \mathds{1}(a_t \leq u_t < b_t)\right\}
	\mathds{1}(a_1 \leq u_1 < b_1)\,,\label{eq:extdensitytsu}
\end{eqnarray}
with $U_1$ marginally uniform on $[0,1]$. These authors employ a D-vine copula, and
show how
 estimation using this extended likelihood 
can be undertaken 
by Bayesian data augmentation, where the values of $\uvec$ are generated 
in an MCMC sampling scheme. Alternatively,
\cite{loaiza2019VB} show how to estimate the copula parameters using variational Bayes 
methods~\citep{blei2017}. These calibrate
tractable approximations to the augmented posterior obtained from
the extended likelihood above. They call this approach ``variational Bayes data augmentation''
(VBDA) and 
show it is faster than MCMC and can be employed for much larger $T$ for many choices of copula. 

\subsection{Implicit time series copulas}
\subsubsection{Decomposition}
In early work, \cite{lambert2002} and \cite{freeswang2005,freeswang2006} suggested
adopting the
implicit copula of an auxiliary stochastic process $\{Z_t\}$. 
In this case, the copula density $c_{1:t}$ has the form
at~\eqref{eq:icoppdf}, so that for $t\geq 2$
\[
c_{1:t}(u_1,\ldots,u_t)=f_{Z_{1:t}}(z_1,\ldots,z_t)/\prod_{s=1}^t f_{Z_s}(z_s).
\]
The conditional density at~\eqref{eq:tscopcondpdf} is therefore
\begin{eqnarray}
	f_{Y_{t+1|1:t}}(y_{t+1}|y_{1},\ldots,y_t) 
	&= &f_{U_{t+1|1:t}}(u_{t+1}|u_1,\ldots,u_t)f_{Y_{t+1}}(y_{t+1})\nonumber\\
&=&\frac{f_{Z_{1:t+1}}(z_1,\ldots,z_{t+1})}
{f_{Z_{1:t}}(z_1,\ldots,z_{t})f_{Z_{t+1}}(z_{t+1})}f_{Y_{t+1}}(y_{t+1})\nonumber\\
&=&f_{Z_{t+1|1:t}}(z_{t+1}|z_1,\ldots,z_{t})
\frac{f_{Y_{t+1}}(y_{t+1})}
{f_{Z_{t+1}}(z_{t+1})}\,.\label{eq:tsu}
\end{eqnarray}

\subsubsection{Stationarity}
A major advantage of an implicit time series copula is that for many processes $\{Z_t\}$, the
densities $f_{Z_{t+1|1:t}}$ and $f_{Z_{t+1}}$ are 
straightforward to compute and simulate from, simplifying
parameter estimation and evaluation of predictive distributions.
It is straightforward to show (e.g. see~\cite{chen2006tscop,smith2015}) that if 
$\{Z_t\}$
is a (strongly) stationary stochastic process, 
then $F_{Z_t}$ is time invariant and $\{U_t\}$ is also stationary
because $U_t=F_{Z_t}(Z_t)$ is a monotonic transformation. 
In addition, if the marginal distribution $F_{Y_t}$ is also time invariant, the process 
$\{Y_t\}$ is also stationary.

\subsubsection{Discrete time series data} 
For an implicit copula, the extended likelihood at~\eqref{eq:extdensityz} 
based on $\bm{Z}=(Z_1,\ldots,Z_T)^\top$ with realization $\zvec=(z_1,\ldots,z_T)^\top$ 
can be used instead of that at~\eqref{eq:extdensitytsu}, which is
\begin{eqnarray*}
f_{Y,Z}(\yvec,\zvec) &= &\prod_{t=2}^T\left\{ \mathds{1}\left(F_{Z_t}^{-1}(a_t) \leq z_t < F_{Z_t}^{-1}(b_t)\right)f_{Z_{t|1:t-1}}(z_t|z_1,\ldots,z_{t-1})\right\}\\
&\times & \mathds{1}\left(F_{Z_1}^{-1}(a_1) \leq z_1 < F_{Z_1}^{-1}(b_1)\right)f_{Z_1}(z_1)\,.
\end{eqnarray*}

\subsubsection{Example: Gaussian autoregression copula}\label{ssec:gautocop}
The simplest implicit time series copulas are those based on stationary Gaussian time series models.
\cite{cario1996} and~\citet[pp.259]{joe97} suggest using a zero mean
stationary autoregression of lag length $p$, so that 
\[
Z_s = \sum_{k=1}^p \rho_k Z_{s-k}+ e_s\,,\mbox{ for } s=1,2,\ldots\,,
\]
with $e_s\sim N(0,\sigma^2)$ an
independent disturbance, and parameters $\{\rho_1,\ldots,\rho_p,\sigma^2\}$. 
Then $\bm{Z}_{1:t}=(Z_1,\ldots,Z_t)^\top \sim N_t(\bm{0},\sigma^2\Sigma_{1:t})$, with  $\sigma^2\Sigma_{1:t}$ the usual full rank autocovariance matrix with $\Sigma^{-1}_{1:t}$ a band $p$ matrix that is a function of $\rhovec=(\rho_1,\ldots,\rho_p)^\top$ only.
 
Therefore, the implicit copula of $\bm{Z}_{1:t}$ is the Gaussian copula
 $C_{\mbox{\tiny Ga}}(\uvec;\Omega_{1:t})$ with the autocorrelation matrix $\Omega_{1:t}=\mbox{diag}(\Sigma_{1:t})^{-1/2}\,
\Sigma_{1:t} \,\mbox{diag}(\Sigma_{1:t})^{-1/2}$. The parameter
$\sigma$ does not feature in $\Omega_{1:t}$ (i.e. it is unidentified in the copula), so that it is sufficient to fix it to an arbitrary
value such as $\sigma^2=1$, as is done here. 
Thus, $\Omega_{1:t}$ is only a function of $\rhovec$, so that $\thetavec=\rhovec$ are the copula parameters.  The marginal distribution $Z_t\sim N(0,\gamma_0)$, with variance $\gamma_0$ computed from $\rhovec$.
Denoting
the density of a standard normal as $\phi(\cdot)$, and that of a $N(\mu,\sigma^2)$ as $\phi_1(\cdot;\mu,\sigma^2)$,  the conditional density
\begin{eqnarray*}
f_{U_{t+1|1:t}}(u_{t+1}|u_1,\ldots,u_{t}) &= &f_{Z_{t+1|1:t}}(z_{t+1}|z_1,\ldots,z_{t})/
f_{Z_{t+1}}(z_{t+1})\\
&= &\phi\left(z_{t+1}-\sum_{k=1}^p \rho_k z_{t-k+1} \right)/\phi_1\left(z_{t+1};0,\gamma_0\right)\,,
\end{eqnarray*}
with $z_{t}=\Phi^{-1}_1(u_{t};0,\gamma_0)$ a $N(0,\gamma_0)$ distribution
function evaluated at $u_t$. (The dependence of this conditional density on $\thetavec$ is tacit here.) Thus, the likelihood of a continuous-valued 
series, or the extended likelihood of a discrete-valued series, can be expressed in terms of the 
copula parameters $\rhovec$ and the marginals $F_{Y_1},\ldots,F_{Y_T}$.  
A variety of estimation methods, including standard maximum likelihood, can then be used to estimate the time series copula parameters.

This copula model extends the stationary autoregression from a marginally Gaussian process to one with
any other marginal distribution. This is why~\cite{cario1996} originally labeled it an ``autoregression-to-anything'' transformation, although these authors did not recognize 
it as a Gaussian copula. Interestingly, even though the auxiliary stochastic process $\{Z_t\}$ is conditionally homoscedastic (i.e. $\mbox{Var}(Z_{t+1}|Z_{1:t})=1$) the process
$\{Y_t\}$ need not be so (i.e. it can be heteroscedastic).
To see this, notice that even when $f_{Y_t}=g$ is time invariant,  the conditional density of $Y_{t+1}|Y_{1:t}$ is
\[
f_{Y_{t+1|1:t}}(y_{t+1}|y_{1},\ldots,y_t)=
\phi\left(z_{t+1}-\sum_{k=1}^p \rho_k z_{t-k+1} \right)
\frac{g(y_{t+1})}{\phi_1\left(z_{t+1};0,\gamma_0\right)}\,.
\]
The second moment of this density is not necessarily
a constant with respect to time, as demonstrated in~\cite{smith+vahey2016}.

The usual measures of serial dependence for an autoregression (e.g. autocorrelation or partial autocorrelation matrices) 
can be computed for $\{Z_t\}$.
Spearman correlations, which are unaffected by the choice
of continuous margin(s) $F_{Y_t}$, provide
 equivalent metrics for $\{Y_t\}$. For example, the Spearman autocorrelation at lag $h$ is 
\[
\rho^S_{h}=\frac{6}{\pi} \mbox{arcsin}\left(\frac{\gamma_h}{2\gamma_0 }\right)\,,
\]
where $\gamma_h\equiv \mbox{Cov}(Z_{t+h},Z_t)$ is the autocovariance at lag $h$ for the auxiliary stochastic process, and is a function of $\rhovec$. Other popular measures of concordance, can also
be computed easily for different values of $h$. 

Last, while the Gaussian autoregression copula---or indeed other Gaussian time series copulas, such as those based on Gaussian processes~\citep{Wilson2010}---produces a flexible family of
time series models, the form of serial dependence is still limited. For example, serial 
dependence is both symmetric and has zero tail dependence, which are properties of the Gaussian copula. This motivates the
construction of more flexible time series copulas, as now discussed.

\subsection{Implicit state space copula}\label{sec:sscop}
A wide array of time series and other statistical models can be written in state space form; see~\cite{durbin2012} for an overview of this extensive class.  
\cite{smithman2018} outline how to construct and estimate the implicit time series copulas of such models,
as is now outlined. 

\subsubsection{The copula}
A nonlinear state space model for $\{Z_t\}$ is given by the observation and transition equations
\begin{eqnarray}
	Z_t|\bm{X}_t=\bm{x}_t &\sim &H_t(z_t|\bm{x}_t;\bm{\theta}) \label{eq:obsn}\\
	\bm{X}_t|\bm{X}_{t-1}=\bm{x}_{t-1} &\sim &K_t(\bm{x}_t|\bm{x}_{t-1};\bm{\theta}).
	\label{eq:trans}
\end{eqnarray}
Here, $H_t$ is the distribution function of
$Z_t$, conditional on an $r$-dimensional state vector $\bm{X}_t$.
The states follow a Markov process, with
conditional distribution function $K_t$. Typically,
tractable parametric distributions are  adopted for $H_t$ and $K_t$,
with the parameters denoted collectively as
$\bm{\theta}$. 

A key requirement in evaluating~\eqref{eq:icopcdf}  and~\eqref{eq:icoppdf} is the computation
of the marginal
distribution and density functions of $Z_t$. Marginalizing over $\bm{X}_t$ gives these as
\begin{eqnarray}
	F_{Z_t}(z_t|\bm{\theta}) &= &\int H_t(z_t|\bm{x}_t;\bm{\theta})f_{X_t}(\bm{x}_t|\bm{\theta})
	\mbox{d}\bm{x}_t \nonumber \\
	f_{Z_t}(z_t|\bm{\theta}) &= &\int h_t(z_t|\bm{x}_t;\bm{\theta})f_{X_t}(\bm{x}_t|\bm{\theta})
	\mbox{d}\bm{x}_t\,,\label{eq:mdist}
\end{eqnarray}
where the dependence on $\bm{\theta}$ is denoted explicitly here. The density
$h_t(z_t|\bm{x}_t;\bm{\theta}) =\frac{d}{d z_t} H_t(z_t|\bm{x}_t;\bm{\theta})$, 
and $f_{X_t}(\bm{x}_t|\bm{\theta})$ is the marginal density of the state
variable $\bm{X}_t$. 
Evaluation of the integrals in~(\ref{eq:mdist}) is straightforward either 
analytically or numerically for many choices of state space model used in practice. 
Note that the quantile function $z_t=F^{-1}_{Z_t}(u_t|\bm{\theta})$ is a function 
of $\thetavec$, which can be computed
quickly using the interpolation
method outlined in~\ref{app:A} when $F_{Z_t}$ is time invariant. 

A more challenging problem is the evaluation of
the numerator in~\eqref{eq:icoppdf}. To compute this,
the state vector $\bm{x}=(\bm{x}_1^\top,\ldots,\bm{x}_T^\top)^\top$ with $Tr$-dimensional joint density $f_X$ needs to be integrated
out, with
\begin{eqnarray*}
	f_Z(\bm{z}|\bm{\theta}) &= &\int f_{Z|X}(\bm{z}|\bm{x},\bm{\theta}) f_X(\bm{x}|\bm{\theta})
	\mbox{d}\bm{x} \\
	&= &\int \prod_{t=1}^T \left\{ h_t(z_t|\bm{x}_t;\bm{\theta}) \right\}
	\prod_{t=2}^T \left\{ k_t(\bm{x}_t|\bm{x}_{t-1};\bm{\theta})\right\} f_{X_1}(\bm{x}_1;\bm{\theta}) \mbox{d}\bm{x}
	\,,
\end{eqnarray*}
where $k_t(\bm{x}_t|\bm{x}_{t-1};\bm{\theta})=\frac{d}{d \bm{x}_t}
K_t(\bm{x}_t|\bm{x}_{t-1};\bm{\theta})$.  While there a number of existing methods in the state space
literature to evaluate $f_Z(\bm{z}|\bm{\theta})$ above,
robust Bayesian MCMC methods that generate the states $\xvec$ are very popular.
The same methods can also be employed
estimate the implicit copula as outlined below.

\subsubsection{Bayesian estimation}
Conditional on the states, a continuous time series copula model likelihood is
\begin{equation}
	f(\bm{y}|\bm{x},\bm{\theta})=
	f_{Z|X}(\bm{z}|\bm{x},\bm{\theta})\prod_{t=1}^T \frac{f_{Y_t}(y_t)}{f_{Z_t}(z_t|\bm{\theta})}=
	\prod_{t=1}^T\left\{
	h_t(z_t|\bm{x}_t;\bm{\theta})\frac{f_{Y_t}(y_t)}{f_{Z_t}(z_t|\bm{\theta})}\right\}
	\,.
	\label{eq:ssclike}
\end{equation}
where all components on the right-hand side of~\eqref{eq:ssclike} are known densities.
Computationally,
it is much easier to work with~\eqref{eq:ssclike}, rather than with the decomposition~\eqref{eq:sklarpdf} and
copula density~\eqref{eq:icoppdf}.
Adopting the prior $\pi_\theta(\bm{\theta})$, Bayesian estimation
and inference of the copula parameters $\thetavec$ can be based on 
the MCMC sampler at Algorithm~\ref{alg:sscop}
below, which produces Monte Carlo draws from the posterior of $\thetavec$ 
augmented
with the latent states $\xvec$.

\begin{algorithm}
	\caption{\em (MCMC sampler for a state space copula and continuous $Y_t$)}
	\label{alg:sscop}
	\begin{itemize}
		\item[1.] Generate from 
		$f(\bm{x}|\bm{\theta},\bm{y})=f(\bm{x}|\bm{\theta},\bm{z})\propto \left( \prod_{t=1}^T h_t(z_t|\bm{x}_t;
		\bm{\theta}) \right) f_X(\bm{x}|\bm{\theta})$ using existing methods
		\item[2.] Generate from $f(\bm{\theta}|\bm{x},\bm{y}) \propto
		\left( \prod_{t=1}^T h_t(z_t|\bm{x}_t;\bm{\theta})/f_{Z_t}(z_t|\bm{\theta}) \right) 
		f_X(\bm{x}|\bm{\theta})
		\pi_{\theta}(\bm{\theta})$
	\end{itemize}
\end{algorithm}

Unlike the states $\xvec$, the values 
$\bm{z}=(z_1,\ldots,z_T)^\top$ are not generated in the 
sampling scheme, but instead
are computed as $z_t=F_{Z_t}^{-1}(u_t|\thetavec)$ for each draw of the parameters $\bm{\theta}$.
Crucially,
Step~1 is exactly
the same as that for the underlying state space model, so that any of the wide range of 
existing procedures for generating $\xvec$ can be employed. 
Step~2 can be undertaken using a
Metropolis-Hastings step, with a proposal based
on a numerical or other approximation to the conditional posterior. In Algorithm~\ref{alg:sscop} the marginal distributions $F_{Y_1},\ldots,F_{Y_T}$ are
assumed known.
It is common to estimate these prior
to estimating the copula parameters~\citep{joe2005}, although joint
estimation of the marginals and copula parameters may also be considered.
In a Bayesian analysis
the prior $\pi_\theta(\bm{\theta})$ reflects any constraints required to 
identify $\thetavec$.

\subsubsection{Example: UCSV implicit copula}\label{sec:ucsvcop}
\cite{smithman2018} constructed the implicit copulas of three specific state space models, and 
estimated their parameters for U.S. inflation between 1954:Q1 and 2013:Q4. These included an unobserved component stochastic volatility (UCSV) model, as is now
outlined. To illustrate, 
it is then applied to the same quarterly U.S. inflation
series used by these authors, but updated to include all observations up to 2020:Q2.
This includes the impact of the Covid-19 pandemic, which this flexible copula time
series model is well-suited to capture.

\subsubsection*{The copula and identifying constraints}
The UCSV model is specified for bivariate state vector $\bm{x}_t=(\mu_t,\zeta_t)^\top$ as
\begin{eqnarray}
	Z_t|\bm{X}_t=\bm{x}_t &\sim &N(\mu_t,\exp(\zeta_t)) \nonumber \\
	\mu_t |\bm{X}_{t-1}=\bm{x}_{t-1} &\sim &N(\bar \mu+\rho_{\mu}(\mu_{t-1}-\bar \mu),\sigma_{\mu}^2) 
	\nonumber \\
	\zeta_t|\bm{X}_{t-1}=\bm{x}_{t-1} &\sim &N(\bar \zeta+\rho_\zeta (\zeta_{t-1}-\bar \zeta),\sigma_\zeta^2)\,. \label{eq:svuc} 
\end{eqnarray}
The parameters $|\rho_\mu|<1$ and $|\rho_\zeta|<1$, which ensures $\{Z_t\}$
is a
(strongly) stationary first order Markov process. The
mean $E(Z_t)=\bar \mu$, which is unidentified in the implicit copula at~\eqref{eq:icopcdf}, and set $\bar \mu=0$ here. 
The marginal variance
$\mbox{Var}(Z_t)=s^2_\mu+\exp(\bar \zeta + s^2_\zeta/2)$,
where
$s^2_\mu=\sigma^2_\mu/(1-\rho_\mu^2)$ and $s^2_\zeta=\sigma^2_\zeta/(1-\rho_\zeta^2)$.
The variance $\mbox{Var}(Z_t)$ is unidentified in the copula, and
setting this equal to one provides an equality constraint 
on $\bar \zeta = \log(1-s^2_\mu)-\frac{s_\zeta^2}{2}$. In addition,
$\exp(\bar \zeta+s^2_\zeta/2)\geq 0$, giving the inequality constraint
$0<\sigma^2_\mu\leq (1-\rho_\mu^2)$.
With these identifying constraints, the 
dependence parameters of the resulting implicit copula are
$\bm{\theta}=\{\rho_\mu,\rho_\zeta,\sigma^2_\mu,\sigma^2_\zeta\}$. 

\subsubsection*{Evaluating the auxiliary margin}
Because $\{Z_t\}$ is stationary, the marginal density $f_{Z_t}$ at~(\ref{eq:mdist}) is time-variant
and given by
\[
f_{Z_1}(z;\bm{\theta})=\int\int \phi_1\left( z;\mu,\exp(\zeta)\right) 
\phi_1(\zeta;\bar \zeta, s_\zeta^2) \phi_1(\mu;0,s^2_\mu)d\mu d\zeta\,.
\] 
The integral in $\mu$ can be recognized as that of a Gaussian density to give
\begin{eqnarray*}
	f_{Z_1}(z;\bm{\theta}) &= &\int \phi_1(z;0,w(\zeta)^2)\phi_1(\zeta;\bar \zeta,s_\zeta^2)d\zeta \\
	F_{Z_1}(z;\bm{\theta}) &= &\int \Phi_1(z;0,w(\zeta)^2)\phi_1(\zeta;\bar \zeta,s_\zeta^2)d\zeta\,,
\end{eqnarray*}
with $w(\zeta)^2=s^2_\mu+\exp(\zeta)$. Computing the (log) copula
density at~(\ref{eq:icoppdf}) requires
evaluating $\log(f_{Z_1})$ and the quantile function $F_{Z_1}^{-1}$ at all $T$
observations. To do so, the accurate and fast numerical method described 
in~\ref{app:A} is used.
\begin{figure}[tbh]	
	\begin{center}
		\includegraphics[width=0.8\textwidth]{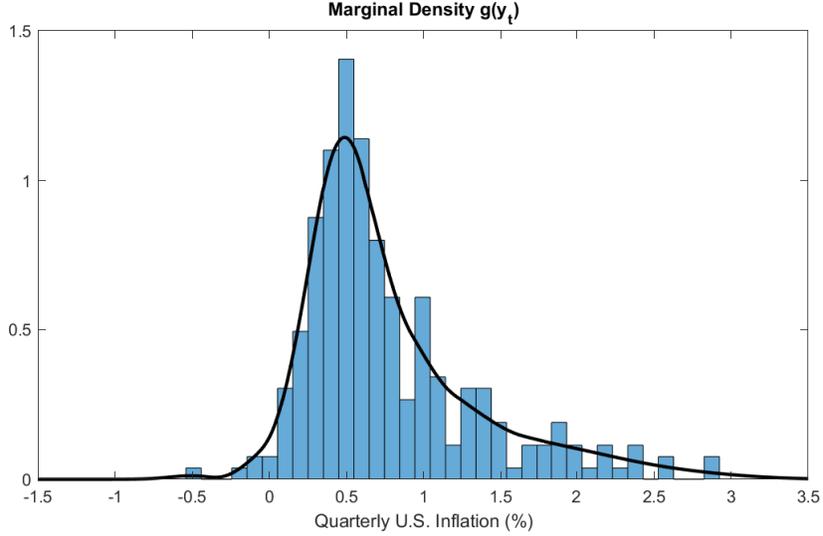}
	\end{center}
	\caption{Estimated Marginal Distribution of U.S. Quarterly Inflation. The histogram is of observations of quarterly U.S. inflation between 1954:Q1 and 2020:Q2
		computed as the quarterly differences of the logarithm of the GDP price deflator. 
		The line is the adaptive KDE estimate of time-invariant marginal $G=F_{Y_t}$.}\label{fig:Ghat}
\end{figure}

\subsubsection*{Copula parameter estimation}
The parameters of this time series copula model are estimated using their Bayesian posterior
with the prior $\pi_{\theta}(\thetavec)\propto \frac{1}{\sigma^2_\mu \sigma^2_\zeta}\mathds{1}(\thetavec \in R_\theta)$,
where $R_\theta$ is the region of parameter values that correspond to the 
constraints outlined above.  Algorithm~\ref{alg:sscop} can be used to estimate the copula parameters, where at~Step 1 
the state vector $\xvec$ is partitioned into
$\bm{\mu}=(\mu_1,\ldots,\mu_T)^\top$ and $\bm{\zeta}=(\zeta_1,\ldots,\zeta_T)^\top$, and
generated using the two separate steps:
\begin{itemize}
	\item[] Step~1a. Generate from $f(\bm{\mu}|\bm{\theta},\bm{\zeta},\bm{y})
	\propto \prod_{t=1}^T \phi_1 \left( z_t;\mu_t,\exp(\zeta_t)\right)f(\bm{\mu}|\bm{\theta})$
	\item[] Step~1b. Generate from $f(\bm{\zeta}|\bm{\theta},\bm{\mu},\bm{y})
	\propto \prod_{t=1}^T \phi_1 \left( z_t;\mu_t,\exp(\zeta_t)\right)f(\bm{\zeta}|\bm{\theta})$
\end{itemize}
The posterior of $\bm{\mu}$ in Step~1a can be recognized as 
normal with zero mean and a band one precision
matrix, so that generation is both straightforward and fast.
There are a number of efficient methods to generate $\bm{\zeta}$ in Step~1b
in the literature, and 
the fast ``precision sampler'' for the latent 
states outlined in~\cite{chan2009} is used here. In Step~2 of the sampler, a normal approximation
is used as a proposal density for the Metropolis-Hastings step, which has high 
acceptance rates in practice.

\begin{figure}[tbh]	
	\begin{center}
		\includegraphics[width=0.8\textwidth]{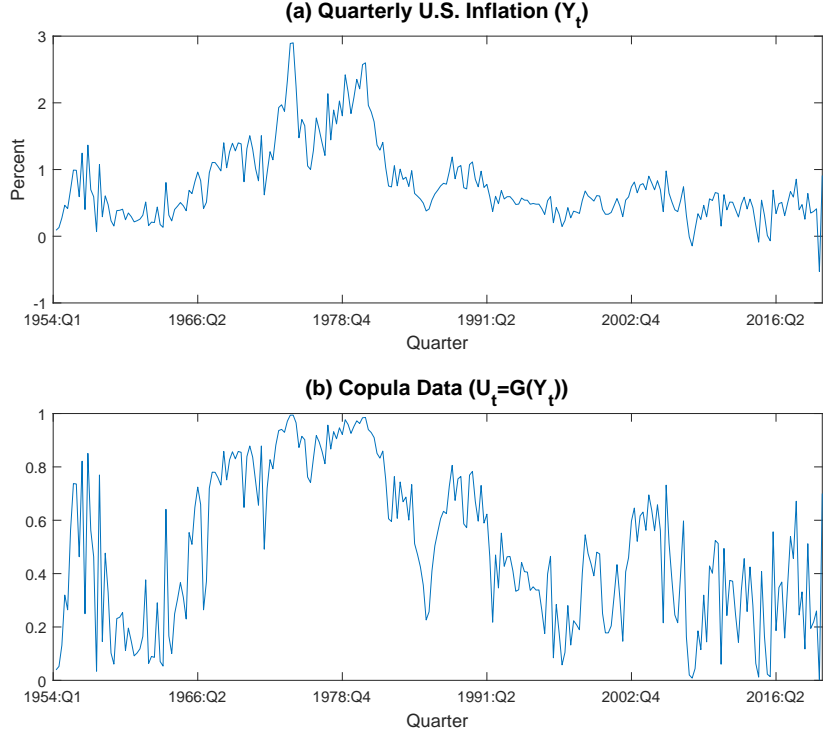}
	\end{center}
	\caption{U.S. Inflation and Copula Data. Panel~(a) plots the $T=266$ observations of U.S. quarterly inflation.
		Panel~(b) plots the corresponding copula data $u_t=G(y_t)$. The impact of the COVID-19 pandemic is seen in the last two observations.}\label{fig:yu}
\end{figure}

\subsubsection*{Empirical results}
The adaptive kernel density estimator (AKDE) of~\cite{shimazaki2010} is used to estimate a time-invariant marginal distribution
$G$ of $Y_t$, and is presented in Figure~\ref{fig:Ghat}. The estimated density is smooth, positively skewed,
and heavy-tailed; it accounts for both high (e.g. $2.9\%$ in 1974:Q3) and low (e.g. $-0.529\%$ in 2020:Q1) values. Figure~\ref{fig:yu} plots the time series, plus the 
copula data $u_t=G(y_t)$ for $t=1,\ldots,T$.

\begin{figure}[tbh]	
	\begin{center}
		\includegraphics[width=0.8\textwidth]{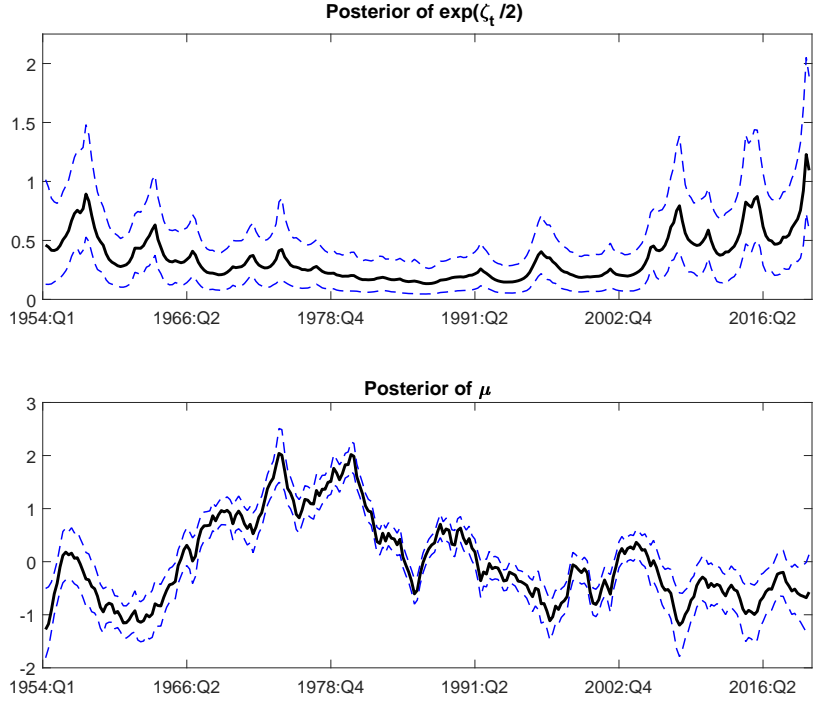}
	\end{center}
	\caption{Bayesian posterior estimates of the latent states. Panel~(a) plots
		$\exp(\zetavec/2)$ and panel~(b) $\muvec$. The posterior mean is the solid line, and the dashed lines are the 5\% and 95\% posterior quantiles.}\label{fig:x}
\end{figure}

To summarize the posterior estimate of the implicit copula, Figure~\ref{fig:x} plots the 
posterior means and 90\% posterior intervals for $\muvec$ and 
$\exp(\zetavec/2)$, which are the mean and standard deviation of the auxiliary 
vector $\bm{Z}$.
While these are not the mean and standard deviation of $\bm{Y}$, they do 
account for movements in the moments of this variable, 
and the impact of the Covid-19 pandemic on 
2020 can be seen as a sharp jump in $\exp(\zeta_t/2)$ in panel~(a), while the inflationary
period of the 1970's can be see in high values of $\mu_t$ in panel~(b).

\begin{figure}[H]	
	\begin{center}
		\includegraphics[width=1\textwidth]{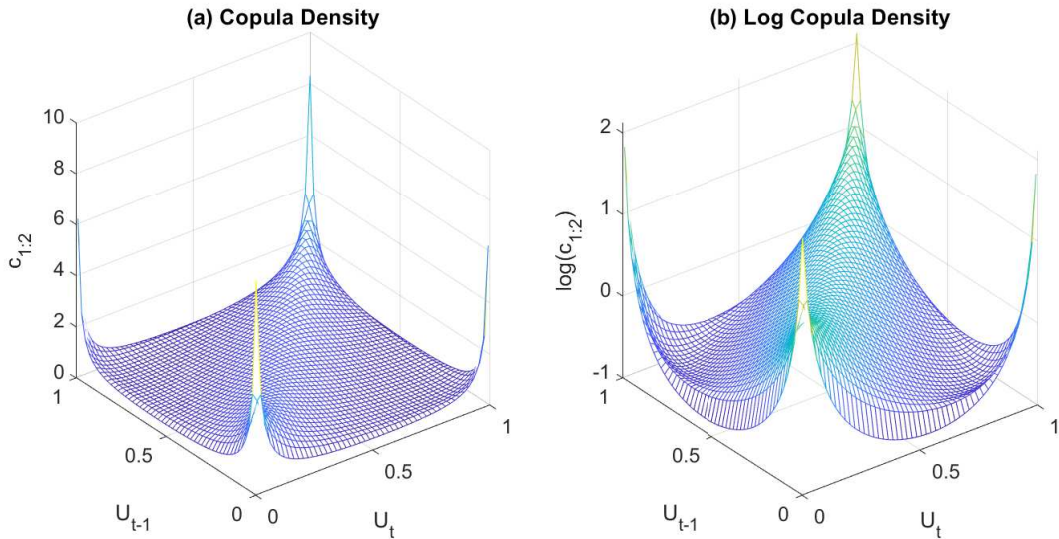}
	\end{center}
	\caption{Panel~(a) bivariate marginal copula density $c_{t-1:t}(u_{t-1},u_{t})$ evaluated at the posterior mean of $\thetavec$ (the values of which are $\hat \rho_\zeta=0.896$, $\hat \sigma_\zeta^2=0.407$, $\hat \rho_\mu=0.960$ and $\hat \sigma_\mu^2=0.059$). 
	Panel~(b) presents the same density on the logarithmic scale. The spikes in the four corners are indicative of serial dependence in the variance
	of the series. The ridge from (0,0) to (1,1) is indicative of serial dependence in the level of the series.}\label{fig:C12}
\end{figure}

There is high serial
dependence in both state variables. One way to show how this affects the time
series copula is to consider the bivariate margin $c_{t-1:t}(u_{t-1},u_t|\thetavec)$ of the copula density, which
is time invariant.
It is given by $c_{1:2}(u_1,u_2|\thetavec)=f_{Z_{1:2}}(z_1,z_2|\thetavec)/f_{Z_1}(z_1|\thetavec)f_{Z_1}(z_2|\thetavec)$, where
the numerator is computed by numerical integration.
Figure~\ref{fig:C12} plots this density at the posterior mean of the copula parameters
$\thetavec$, and two interesting features can be seen. First, ``spikes'' at the corners (i.e. near (0,0), (0,1), (1,0) and (1,1)) are indicative of strong dependence in the {\em volatility} of the series; see~\cite{loaiza2018hetero} and~\cite{bladt2021} for a discussion of such a pattern in a time series copula. 
Second, the positive ``ridge'' running from (0,0) to (1,1) is indicative of positive dependence in the {\em level} of the series. Both these features
are well-known aspects of inflation time series, and the implicit copula captures them both while also
allowing for the asymmetric marginal distribution in Figure~\ref{fig:Ghat}. The
time series copula model therefore allows for more realistic modeling of tail risk than
the standard UCSV model, and improves density forecast accuracy, including in the tails.  

%% file: sec5.tex
\section{Implicit copulas for multivariate time series}\label{sec:mtimeseries}
Copulas have been used extensively to capture cross-sectional dependence 
in a multivariate stochastic 
process $\{\bm{Y}_t\}$, where $\bm{Y}_t=(Y_{1,t},\ldots,Y_{d,t})^\top$; see~\cite{patton2006}, \cite{rodriguez2007}, \cite{hafner2012} 
and~\cite{creal2015} for just some examples.
These models typically capture serial dependence through existing 
marginal time series models; for example,
heteroscedastic models are normally used for financial returns. 
An alternative is to use a single high-dimensional---but parsimonious---copula to 
capture both serial and cross-sectional dependence jointly.
An advantage of this approach is that the marginal
distribution of each variable can be modeled directly, including
as non-parametric. It is this type
of time series copula that is the focus of this section. 

\subsection{Multivariate time series copula models}
\subsubsection{Copula model}
If the random vector $\bm{Y}=(\bm{Y}_1^\top,\ldots,\bm{Y}_T^\top)^\top$,
then
$F_Y$ is given by~\eqref{eq:sklarcdf} with $m=Td$ and the order of the elements of $\bm{Y}$
 determines the interpretation of $C$. 
If all variables are continuous,
$f_Y$ is given by~\eqref{eq:sklarpdf}, so that
\begin{equation}
f_Y(\yvec)=c(\bm{u})\prod_{j=1}^d \prod_{t=1}^T f_{Y_{j,t}}(y_{j,t})\,,\label{eq:mtscopdens}
\end{equation}
where $\yvec=(\yvec_1^\top,\ldots,\yvec_T^\top)^\top$,
$\yvec_t=(y_{1,t},\ldots,y_{d,t})^\top$, $\uvec=(\uvec_1^\top,\ldots,\uvec_T^\top)^\top$
and $\uvec_t=(u_{1,t},\ldots,u_{d,t})^\top$. For
discrete-valued variables the mass function is given by~\eqref{eq:copmass}, and an extended likelihood
for $(\bm{Y},\bm{U})$ is given by~\eqref{eq:extdensity}; see~\cite{loaiza2019VB}.
Marginal models for each of the $d$ series
are required, and one option is to assume they are
time-invariant with distribution functions $G_1,\ldots,G_d$, which can be estimated separately.

\subsubsection{Copula choice}
Selecting an appropriate $Td$-dimensional copula with density $c$ at~\eqref{eq:mtscopdens} 
is difficult because it needs to capture three forms of 
dependence: (i)~cross-sectional contemporaneous, (ii)~within-series serial, and (iii)~cross-series serial. 
One solution is to use a vine copula; for example
see~\cite{brechmann2015} for the two-dimensional case,~\cite{smith2015} and~\cite{loaiza2018hetero} for 
D-vine copulas, \cite{beare2015} for an M-vine and~\cite{zhao2020} for an alternative
vine-based copula; see also~\cite{remillard2012,nagler2020}.
However, 
implicit copulas constructed from existing multivariate time series
models offer a tractable alternative to vines, particularly for series where $T$ and/or $d$ are 
large.

\subsection{Gaussian vector autoregression copula}
The most popular implicit copula for multivariate time series is that of
a Gaussian vector autoregression (VAR) for $\{\bm{Z}_t\}$. This is an extension of
the autoregression copula in Section~\ref{ssec:gautocop}. 
Consider the following VAR with lag $p$, 
\begin{equation}
\bm{Z}_t= \sum_{j=1}^p B_j \bm{Z}_{t-j} + \evec_t\,, \mbox{ where }\evec_t \sim N(\bm{0},\Sigma)\,.\label{eq:var}
\end{equation}
The mean is set to zero because it is unidentified in the copula, and 
the variances are fixed so that $\mbox{Var}(Z_{j,t})=1$.
Then $\bm{Z}=(\bm{Z}_1^\top,\ldots,
\bm{Z}_T^\top)^\top\sim N_{Td}(\bm{0},\Omega)$, where $\Omega$ is the block Toeplitz
correlation matrix of this process. This is a matrix of $(T\times T)$ blocks, with the
$(s,t)$th block being given by $\Omega_h \equiv \mbox{Corr}(\bm{Z}_{t+h},\bm{Z}_t)$ for 
$h=|t-s|$ and $t\geq s$; for example, see~\citet[p.30]{lutkepohl2005}. Because the Gaussian copula
is closed under marginalization, the $d$-dimensional marginal distribution in $\bm{Y}_t$ also has
a Gaussian copula function $C_{\mbox{\tiny Ga}}(\uvec_t;\Omega_0)$. For example, for 
continuous data $f_{Y_t}(\yvec_t)=c_{\mbox{\tiny Ga}}(\uvec_t;\Omega_0)\prod_{j=1}^d f_{Y_{j,t}}(y_{j,t})$.

For continuous time series, a 
straightforward approach to estimate the model is to first estimate
appropriate marginal distributions $F_{Y_{j,t}}$; for example, by assuming time-invariance
in the marginals and applying a kernel density estimator to each of the $d$ series. 
Second, compute
the auxiliary data $z_{j,t}=\Phi^{-1}(F_{Y_{j,t}}(y_{j,t}))$ for $i=1,\ldots,d$ and 
$t=1,\ldots,T$. Third, apply standard likelihood-based methods for Gaussian VARs
directly to this auxiliary data to estimate the unknown parameters $B_1,\ldots,B_p,\Sigma$.
From these $\Omega$ can be computed, although this can be impractical to evaluate when $m=Td$ is large
and there is often no need to do so.

The conditional density for $\bm{Y}_{t+1}|\bm{Y}_{1:t}$ can be derived in a similar manner 
as for the univariate autoregression copula model. 
This is given by
\begin{eqnarray*}
	f_{Y_{t+1|1:t}}(\yvec_{t+1}|\yvec_1,\ldots,\yvec_{t}) 
	&=&f_{Z_{t+1|1:t}}(\zvec_{t+1}|\zvec_1,\ldots,\zvec_t)\frac{f_{Y_{t+1}}(\yvec_{t+1})}
	{f_{Z_{t+1}}(\zvec_{t+1})}\\
	&=&\phi_d\left(\zvec_{t+1};\sum_{j=1}^p B_j \bm{Z}_{t-j},\Sigma\right)
	\frac{c_{\mbox{\tiny Ga}}(\bm{u}_{t+1};\Omega_0)\prod_{j=1}^d g_j(y_{j,t+1})}
	{\phi_d(\zvec_{t+1};\bm{0},\Omega_0)}
\end{eqnarray*}
where time-invariant marginal densities $g_1,\ldots,g_d$ are assumed.
Drawing from this conditional
distribution is straightforward by first simulating
$\bm{Z}_{t+1}$ directly
from~\eqref{eq:var}, and then transforming to a draw $\bm{Y}_{t+1}=(G_1^{-1}(\Phi(Z_{1,t+1})),\ldots,G_d^{-1}(\Phi(Z_{d,t+1})))^\top$, which 
can be used to compute the predictive distribution. 

\subsubsection{Further reading}
\cite{biller2003} were the first to construct the Gaussian VAR copula via transformation, 
but did not recognize it as a Gaussian copula and called it
a ``Vector-Autoregressive-To-Anything'' distribution. \cite{smith2015} and~\cite{smith+vahey2016} 
also consider this Gaussian copula model, its D-vine 
representation 
and apply
it to multivariate macroeconomic and financial forecasting. Similar 
to the univariate case, the Gaussian VAR copula model can capture a degree of 
heteroscedasticity in the time series given a suitable choice of marginal distributions
$G_1,\ldots,G_d$ for the $d$ time series; see~\cite{smith+vahey2016} for a demonstration. 
There is also a growing interest in multivariate times series copulas in 
machine learning. For example,~\cite{salinas2019high} construct a Gaussian copula
from low rank factor decomposition where the small number of factors follow a 
Gaussian process with recurrent neural network (RNN) dynamics. \cite{klein2020deep}
propose constructing a Gaussian copula process that is constructed as
the implicit copula of 
an RNN with Gaussian errors. 

Existing econometric applications of multivariate time series 
often have parameters that vary over time (widely called a ``dynamic'' model) along with
substantial regularization; see~\cite{bitto2019}, \cite{huber2020} and references
therein. These features can also be employed for the parameters of implicit copulas. For example,
\cite{smith+vahey2016} use Bayesian selection on the D-vine representation of the Gaussian VAR copula 
for regularization, \cite{creal2015}, \cite{oh2017} and~\cite{opschoor2020} allow the 
parameters of elliptical copulas to vary over time, and
\cite{oh2020dynamic} consider a dynamic skew $t$ copula. In another approach 
\cite{loaiza2018hetero}  
extend the UCSV model in Section~\ref{sec:ucsvcop} to the multivariate case and
show how to construct its implicit copula. In all these studies, the copula models are more
accurate than non-copula benchmarks, and the implicit copulas used are scalable to high dimensions.

%% file: sec6.tex
\section{Regression copula processes}\label{sec:regcopprocess}
Copula models with regression margins have been used widely; for examples, see~\cite{pitt2006}, \cite{song2009}, \cite{masarotto2012} and
\cite{klein2016}.
However, another usage of a copula with regression data is to capture the dependence between multiple observations
on a single dependent variable $Y$, conditional on the covariate values. 
This defines a copula process~\citep{Wilson2010} on the covariate
space, which~\cite{smith+k19} call a ``regression copula''. When combined with a flexible marginal distribution for $Y$, it
specifies a new distributional regression model. This is where
the covariates affect the entire distribution of $Y$.
\cite{KleSmi2019} and~\cite{smith+k19} consider a regression copula 
that is the implicit copula 
of the joint distribution of observations in an auxiliary regression model. 
They are inherently high dimensional, yet can be estimated in reasonable time 
using Bayesian methods.
The idea is outlined in this section for continuous $Y$, 
and greater detail can be found in these papers.

\subsection{The basic idea of a regression copula}

\subsubsection{The copula process model}
Consider $N>1$ realizations $\bm{Y}_{1:N}=(Y_1,\ldots,Y_N)^\top$ of a dependent variable
with corresponding values $\xvec_{1:N}=\{\xvec_1,\ldots,\xvec_N\}$ for
$p$ covariates, with $\xvec_i=(x_{i,1},\ldots,x_{i,p})^\top$. Then application of
Sklar's theorem to
the distribution of $\bm{Y}_{1:N}|\xvec_{1:N}$  gives
\[
F_{Y_{1:N}}(\yvec_{1:N}|\xvec_{1:N})=C^\dagger_{1:N}\left(F_{Y_1}(y_1|\xvec_1),\ldots,
F_{Y_N}(y_N|\xvec_{N})\,;\,\xvec_{1:N}\right)\,.
\] 
The $N$-dimensional copula function $C^\dagger_{1:N}(\cdot\,;\,\xvec_{1:N})$ is a copula process 
on the covariate space, and $F_{Y_i}(y_i|\xvec_i)$ is the distribution function of $Y_i|\xvec_i$. 
Both are typically unknown, and in a copula model these are selected to define
the distribution.
One tractable but effective simplification 
is to allow the covariates to only affect the dependent variable through 
the copula function, so that $Y_i$ is 
{\em marginally} independent
of $\xvec_i$. In this case, 
\begin{equation}
F_{Y_{1:N}}(\yvec_{1:N}|\xvec_{1:N})=C_{1:N}\left(F_{Y_1}(y_1),\ldots,
F_{Y_N}(y_N)\,;\,\xvec_{1:N},\thetavec \right)\,,\label{eq:copregcdf}
\end{equation}
with $\thetavec$ unknown copula parameters that are unaffected by the dimension $N$ and require estimation. Here, the {\em joint} distribution of $\bm{Y}_{1:N}$ is dependent on $\xvec_{1:N}$ via the copula,
so that the conditional distribution $Y_{N}|(\bm{Y}_{1:N-1}=\yvec_{1:N-1}),\xvec_{1:N}$ is also. The latter is employed as the predictive distribution of the regression 
model, as discussed
further below.
 
When the dependent variable is continuous, the joint density is
\begin{equation}
f_{Y_{1:N}}(\yvec_{1:N}|\xvec_{1:N})=c_{1:N}(u_1,\ldots,u_N\,;\,\xvec_{1:N},\thetavec)
\prod_{i=1}^N f_{Y_i}(y_i)\,, \label{eq:copregpdf}
\end{equation}
with $u_i=F_{Y_i}(y_i)$.
An advantage (that is also in common with
the time series copula models discussed in Section~\ref{sec:timeseries}) is that 
if $F_{Y_i}(y_i)\equiv G(y_i)$ is assumed to be invariant with respect to the index $i$, then 
$G$ can be estimated using non-parametric or other flexible estimators.
The remaining component of the copula model at~\eqref{eq:copregcdf} is the choice of copula process,
which is aptly called a regression copula
because it is a function of $\xvec_{1:N}$.

\subsubsection{Distributional regression}
To see how~\eqref{eq:copregpdf} defines
a distributional regression model, consider the 
predictive density for a continuous-valued dependent variable.
For a sample of size
$n$ with covariate values $\xvec_{1:n}$ and dependent variable values $\bm{Y}_{1:n}=\yvec_{1:n}$
 arising from~\eqref{eq:copregpdf},
 the predictive distribution of the subsequent value $Y_{n+1}$ with observed 
 covariates $\xvec_{n+1}$ is defined to be that
of $Y_{n+1}|\xvec_{1:n+1},\yvec_{1:n}$, which has density
\begin{eqnarray}
f_{\mbox{\tiny pred}}(y_{n+1}|\xvec_{n+1},\thetavec) &\equiv& 
f(y_{n+1}|\xvec_{1:n+1},\yvec_{1:n},\thetavec) =
\frac{f(\yvec_{1:n+1}|\xvec_{1:n+1},\thetavec)}
{f(\yvec_{1:n}|\xvec_{1:n},\thetavec)} \nonumber\\
&= &\frac{c_{1:n+1}(u_1,\ldots,u_{n+1};\xvec_{1:n+1},\thetavec)}
{c_{1:n}(u_1,\ldots,u_n;\xvec_{1:n},\thetavec)}f_{Y_{n+1}}(y_{n+1})\nonumber\\
& =&f(u_{n+1}|\uvec_{1:n},\xvec_{1:n+1},\thetavec)f_{Y_{n+1}}(y_{n+1})\,. \label{eq:predyireg}
\end{eqnarray}
Thus, the predictive density is a function of the covariate vector $\xvec_{n+1}$,  as well as those of the 
sample $\xvec_{1:n}$. Moreover, the entire distribution (not just the first or other moments of $Y_{n+1}$) is a function 
of  $\xvec_{n+1}$ as illustrated empirically in Section~\ref{ssec:assetpricing}.

\subsection{Implicit regression copula process}\label{ssec:iregcopproc}
One regression copula process $C_{1:N}$ that can be used
at~\eqref{eq:copregcdf} is an implicit copula derived from an existing regression model, as now discussed.
  
\subsubsection{The copula}
Implicit regression copulas are constructed as 
in Section~\ref{sec:implicitcops}, but  when also conditioning on the covariate values; i.e. from an ``auxiliary regression''
model. Consider a regression model for the auxiliary vector $\bm{Z}_{1:N}=(Z_1,\ldots,Z_N)^\top$ with 
covariate values
$\xvec_{1:N}$ and parameter vector $\thetavec$. Denote the joint
distribution function of $\bm{Z}_{1:N}|\xvec_{1:N},\thetavec$ as $F_{Z_{1:N}}(\cdot|\xvec_{1:N},\thetavec)$, with $i$th marginal
$F_{Z_i}(\cdot|\xvec_i,\thetavec)$.  Then, extending the definition in Table~\ref{tab:icop}, the following transformations define a regression 
copula model
\[
U_i=F_{Z_i}(Z_i|\xvec_i,\thetavec)\,,\mbox{  and } Y_i=F_{Y_i}^{-1}(U_i)\,.
\]
If $\zvec_{1:N}=(z_1,\ldots,z_N)^\top$, $z_i=F_{Z_i}^{-1}(u_i|\xvec_i,\thetavec)$ 
and $m=N$, then the implicit copula function at~\eqref{eq:icopcdf} and density at~\eqref{eq:icoppdf} for this model are given by
\begin{eqnarray}
	 	C_{Z_{1:N}}(\uvec_{1:N};\xvec_{1:N},\thetavec) &\equiv &F_{Z_{1:N}}\left(F_{Z_1}^{-1}(u_1|\xvec_1,\thetavec),\ldots,F_{Z_N}^{-1}(u_N|\xvec_N,\thetavec)|\xvec_{1:N},\thetavec\right)\,,\label{eq:iregcopcdf}\\
c_{Z_{1:N}}(\uvec_{1:N};\xvec_{1:N},\thetavec) &\equiv &\frac{f_{Z_{1:N}}(\zvec_{1:N}|\xvec_{1:N},\thetavec)}{\prod_{i=1}^N f_{Z_i}(z_i|\xvec_i,\thetavec)}\,.\label{eq:iregcoppdf}
\end{eqnarray}
In~\eqref{eq:iregcoppdf} $f_{Z_i}(z_i|\xvec_i,\thetavec)$ is the density function of the auxiliary variable $Z_i$, conditional on
the covariates $\xvec_i$. 
These expressions for $C_{Z_{1:N}}$ and $c_{Z_{1:N}}$
can then be used 
in~\eqref{eq:copregcdf} and~\eqref{eq:copregpdf} to specify a distributional regression.

If in the auxiliary regression
$f_{Z_{1:N}}(\zvec_{1:N}|\xvec_{1:N},\thetavec)=\prod_{i=1}^N f_{Z_i}(z_i|\xvec_i,\thetavec)$, then 
from~\eqref{eq:iregcoppdf} the implicit copula is the trivial independence copula. 
Thus, only distributions where $\bm{Z}_{1:N}$ are dependent are useful for 
constructing an implicit regression copula, as in Section~\ref{sec:lrc} below.

\subsubsection{Predictive density}
Employing the copula density at~\eqref{eq:iregcoppdf} for that in the 
predictive density at~\eqref{eq:predyireg}, gives the following:
\begin{eqnarray}
f_{\mbox{\tiny pred}}(y_{n+1}|\xvec_{n+1},\thetavec)& =& 
\frac{f_{Z_{1:n+1}}(\zvec_{1:n+1}|\xvec_{1:n+1},\thetavec)}{f_{Z_{1:n}}(\zvec_{1:n}|\xvec_{1:n},\thetavec)f_{Z_{n+1}}(z_{n+1}|\xvec_{n+1},\thetavec)}
f_{Y_{n+1}}(y_{n+1})\nonumber\\
&= &f_{Z_{n+1|1:n}}(z_{n+1}|\zvec_{1:n},\xvec_{1:n+1},\thetavec)\frac{f_{Y_{n+1}}(y_{n+1})}{f_{Z_{n+1}}(z_{n+1}|\xvec_{n+1},\thetavec)}\,.\label{eq:predireg}
\end{eqnarray}
To evaluate~\eqref{eq:predireg} in practice, a point estimate of $\thetavec$ can
be used. In a Bayesian analysis another option exists, where $\thetavec$ is
integrated out with respect to its posterior density $f(\thetavec|\yvec)$ to obtain
\[
f_{\mbox{\tiny pred}}^{\mbox{\tiny Bayes}}(y_{n+1}|\xvec_{n+1})=
\int f_{\mbox{\tiny pred}}(y_{n+1}|\xvec_{n+1},\thetavec)f(\thetavec|\yvec)\mbox{d}\thetavec\,.
\] 
This is called the ``posterior predictive density'', and 
evaluation of the integral 
is usually undertaken using draws
obtained from an MCMC sampling scheme.

\subsection{Linear regression copula}\label{sec:lrc}
In principle, implicit copula processes outlined above 
can be constructed from a wide range of different regression models. 
\cite{KleSmi2019} suggest doing so for a Gaussian linear regression, as now outlined.

\subsubsection{The copula}
For a dependent variable $\widetilde{Z}_i$,
consider the linear regression 
\[
\widetilde{Z}_i=\xvec_i^\top \betavec + \sigma e_i\,,
\] 
with $e_i$ distributed independently $N(0,1)$. 
Conditional on both $\xvec_i$ and the parameters $\betavec,\sigma^2$, the elements of
$\widetilde{\bm{Z}}_{1:N}=(\widetilde{Z}_1,\ldots,\widetilde{Z}_N)^\top$ are distributed
independently,
so that their joint distribution cannot
be used directly to specify a useful regression copula with density at~\eqref{eq:iregcoppdf}. 
However, a Bayesian framework can be employed where $\betavec$ is treated as random and marginalized out of the 
distribution for $\widetilde{\bm{Z}}_{1:N}$, the elements of which are then dependent.  From this distribution a useful implicit
regression copula can be formed as below. 

If $B =[\xvec_1|\xvec_2|\cdots|\xvec_N]^\top$
is the  $(N \times p)$ regression design matrix, then the regression can
be written as the linear model
\begin{equation}
	\widetilde{\bm{Z}}_{1:N}|\xvec_{1:N},\betavec,\sigma^2 \sim N(B \betavec,\sigma^2 I). \label{eq:linearreg}
\end{equation}
The conjugate proper prior
\begin{equation}
	\betavec|\sigma^2\sim N(\bm{0},\sigma^2 P(\thetavec)^{-1})\,,\label{eq:conjprior}
\end{equation} 
is used, where 
the precision matrix $P(\thetavec)$ is of full rank $p$ and a function of
$\thetavec$. 
It is necessary to 
assume a proper prior for $\betavec$, because it ensures that the distribution 
with $\betavec$ integrated out is also proper. Doing so (by recognizing a normal in $\betavec$) gives
\begin{equation}
	\widetilde{\bm{Z}}_{1:N}|\xvec_{1:N},\thetavec,\sigma^2 \sim N\left(\bm{0},\sigma^2(I - B  \Sigma B ^\top)^{-1}\right)\,, \label{eq:reducednorm}
\end{equation}
with $\Sigma=(B ^\top B  + P(\thetavec))^{-1}$. Application of the Woodbury formula further
simplifies the variance matrix at~\eqref{eq:reducednorm} as
\[
\sigma^2(I - B \Sigma  B ^\top)^{-1}=\sigma^2 \left(I+B  P(\thetavec)^{-1} B ^\top \right)\,.
\]
The variance of an individual observation $i$ is the $i$th leading diagonal element of this matrix, so that 
$\mbox{Var}(\widetilde{Z}_{i}|\xvec_i,\thetavec,\sigma^2) =\sigma^2(1+\xvec_i^\top P(\thetavec)^{-1} \xvec_i)$. 

The copula of any normal distribution is the Gaussian copula
discussed in Section~\ref{sec:elliptical}. 
The parameter matrix $R$ is the correlation matrix of~\eqref{eq:reducednorm}, 
and it is obtained 
by standardizing $\widetilde{Z}_i$ to have unit variance as follows. Let $s_i=(1+\xvec_i^\top P(\thetavec)^{-1} \xvec_i)^{-1/2}$, 
then define the auxiliary variable of the implicit copula as $Z_i=\frac{s_i}{\sigma}\widetilde{Z}_i$. Thus, if the diagonal matrix
$S(\xvec_{1:N},\thetavec)=\mbox{diag}(s_1,\ldots,s_N)$, then from~\eqref{eq:reducednorm}
the conditional distribution of $\bm{Z}_{1:N}=(Z_1,\ldots,Z_N)^\top$ is 
$\bm{Z}_{1:N}|\xvec_{1:N},\thetavec,\sigma^2 \sim N\left(\bm{0},R\right)$ with correlation matrix
\begin{equation}
	R(\xvec_{1:N},\thetavec)=S(\xvec_{1:N},\thetavec)\left(I+B  P(\thetavec)^{-1} B ^\top \right)S(\xvec_{1:N},\thetavec)\,, \label{eq:Rxtheta}
\end{equation}
and has  
copula function $C_{\mbox{\tiny Ga}}(\uvec;R)$. This is a copula process
on the covariate space because $R$ is a function of the covariate vector $\xvec_{1:N}$ (the notation $R$ and $R(\xvec_{1:N},\thetavec)$ is used interchangeably here.)
The parameter $\sigma^2$ does not feature in $R$ and is unidentified in the copula,
so that $\sigma^2=1$ can be assumed throughout. 

\subsubsection*{Example: horseshoe regularization}
Different implicit copulas can be constructed by using different conditionally Gaussian priors for $\betavec$ at~\eqref{eq:conjprior}.
\cite{KleSmi2019} explore three different choices, including the horseshoe prior of~\cite{CarPol2010} which
is outlined here. This prior
provides regularization of $\betavec$ in the auxiliary regression. 
The prior is given by
\begin{eqnarray*}
\betavec|\lambdavec,\tau &\sim &N\left(\bm{0},\mbox{diag}(\lambdavec)^2\right)\,,\;
\lambdavec=(\lambda_1,\ldots,\lambda_p)^\top\,,\\
\lambda_j|\tau &\sim &\mbox{Half-Cauchy}(0,\tau)\,,\mbox{ and }
\tau \sim \mbox{Half-Cauchy}(0,1)\,,
\end{eqnarray*}
see~\cite{polsonscott12}. The hyper-parameters
of this prior are the parameters of the implicit copula
$\thetavec=(\lambdavec^\top,\tau)^\top$, while the precision matrix $P(\thetavec)=\mbox{diag}(\lambdavec)^{-2}$ is diagonal. 

\subsubsection{Estimation}
For a sample of $n$ observations, 
from~\eqref{eq:copregpdf} and~\eqref{eq:iregcoppdf}, the likelihood is
\[
f_{Y_{1:n}}(\yvec_{1:n}|\xvec_{1:n},\thetavec)=f_{Z_{1:n}}(\zvec_{1:n}|\xvec_{1:n},\thetavec)\prod_{i=1}^n \frac{f_{Y_i}(y_i)}{f_{Z_i}(z_i|\xvec_i,\thetavec)} = \phi_n\left(\zvec_{1:n};\bm{0},R(\xvec_{1:n},\thetavec)\right)
\prod_{i=1}^n \frac{g(y_i)}{\phi(z_i)}\,,
\]
for an invariant marginal distribution with density $f_{Y_i}=g$.
However, even though the likelihood is available in closed form, for large $n$ evaluating and inverting
the $(n\times n)$ matrix $R(\xvec_{1:n},\thetavec)$ to compute the likelihood is computationally demanding.

Instead, it is more efficient to use the
likelihood also conditional on $\betavec$, and integrate out $\betavec$ using an MCMC scheme. (It is stressed here that doing so does not change the implicit copula specification.) 
First, note that
from~\eqref{eq:linearreg} when also conditioning on $\betavec$ the vector $\bm{Z}_{1:n}=S\widetilde{\bm{Z}}_{1:n}\sim N(SB\betavec,S^2)$ (with $\sigma^2=1$ and 
$S=S(\xvec_{1:n},\thetavec)$). 
Also, the Jacobian of the transformation from $\bm{Z}_{1:n}$ to $\bm{Y}_{1:n}$ is 
$J_{Z_{1:n}\rightarrow Y_{1:n}}=\prod_{i=1}^n g(y_i)/\phi(z_i)$. Then, by a 
change of variables, the likelihood also conditional on $\betavec$ is
\begin{equation}
f(\yvec_{1:n}|\xvec_{1:n},\betavec,\thetavec) = f(\zvec_{1:n}|\xvec_{1:n},\betavec,\thetavec)J_{Z_{1:n}\rightarrow Y_{1:n}} =\phi_n(\zvec_{1:n};SB\betavec,S^2)
\prod_{i=1}^n \frac{g(y_i)}{\phi(z_i)}\,,\label{eq:regclike}
\end{equation}
which can be evaluated in $O(n)$ operations because $S$ is a diagonal matrix.
A Bayesian approach that employs this conditional likelihood, evaluates the 
augmented posterior
$f(\betavec,\thetavec|\yvec_{1:n})$ using the sampler at~\ref{alg:regcop}. 
Implementation details for this sampler are given in~\cite{KleSmi2019}.

\begin{algorithm}
	\doublespacing
	\caption{\em (MCMC sampler for regression copula)}
	\label{alg:regcop}
	\begin{itemize}
		\item[1.] Generate from $\betavec|\xvec_{1:n},\thetavec,\yvec_{1:n}$ (which is a Gaussian distribution)
		\item[2.] Generate from $\thetavec|\xvec_{1:n},\betavec,\yvec_{1:n}$ 
	\end{itemize}
\end{algorithm}

%

\subsubsection{Prediction}\label{ssec:predy}
One way to compute the predictive density
that avoids computing $R(\xvec_{1:n},\thetavec)$ or its inverse (and is therefore faster than alternatives), is to also condition on $\betavec$. 
By a change of variables from $Y_{n+1}$ to $Z_{n+1}$,
\begin{eqnarray}
\lefteqn{f(y_{n+1}|\xvec_{1:n+1},\betavec,\thetavec) = f(z_{n+1}|\xvec_{1:n+1},\betavec,\thetavec)
\frac{f_{Y_{n+1}}(y_{n+1})}{\phi(z_{n+1})} }\nonumber\\
& & =
\phi_1\left(z_{n+1};s_{n+1}\xvec_{n+1}^\top \betavec,s_{n+1}^2
\right)\frac{f_{Y_{n+1}}(y_{n+1})}{\phi\left(z_{n+1}\right)}\,,\label{eq:approxregpred}
\end{eqnarray}
where $s_{n+1}=(1+\xvec_{n+1}^\top P(\thetavec)^{-1} \xvec_{n+1})^{-1/2}$ and
$z_{n+1}=\Phi^{-1}(F_{Y_{n+1}}(y_{n+1}))$.

The draws for $\betavec,\thetavec$ from Algorithm~\ref{alg:regcop} can be used
to either integrate out $\betavec,\thetavec$ with respect to the augmented posterior, 
or to compute plug-in point estimates for $\betavec$ and
also $s_{n+1}$. 
If $F_{Y_{n+1}}$ is fixed to its estimate, and $\{\betavec^{[1]},\thetavec^{[1]},\ldots,\betavec^{[J]},\thetavec^{[J]}\}$ are the 
Monte Carlo draws, then the two Bayesian posterior estimators for the predictive 
density are:
\begin{eqnarray*}
	\hat{f}^{\mbox{\tiny Bayes}}_{\mbox{\tiny pred}}(y_{n+1}|\xvec_{n+1})&\equiv& \frac{1}{J}\sum_{j=1}^J  
	f(y_{n+1}|\xvec_{1:n+1},\betavec^{[j]},\thetavec^{[j]})\,, \\
	\hat{f}^{\mbox{\tiny Point}}_{\mbox{\tiny pred}}(y_{n+1}|\xvec_{n+1})&\equiv&   
\phi_1\left(z_{n+1};\hat{s}_{n+1}\xvec_{n+1}^\top \hat{\betavec},\hat{s}_{n+1}^2
\right)\frac{f_{Y_{n+1}}(y_{n+1})}{\phi\left(z_{n+1}\right)}\,, 	
\end{eqnarray*}
with
\[
\hat{\betavec} = \frac{1}{J}\sum_{j=1}^J \betavec^{[j]}\,,\mbox{ and }
\hat{s}_{n+1} = \frac{1}{J}\sum_{j=1}^J \left(1+\xvec_{n+1}^\top P(\thetavec^{[j]})^{-1} \xvec_{n+1}\right)^{-1/2}\,.
\]
In their empirical work,~\cite{KleSmi2019} and~\cite{smith+k19} found that estimates
from these two
estimators were very similar.

\subsubsection{Empirical application: a non-Gaussian asset pricing model}\label{ssec:assetpricing}
\begin{figure}[t]	
	\begin{center}
		\includegraphics[width=0.8\textwidth]{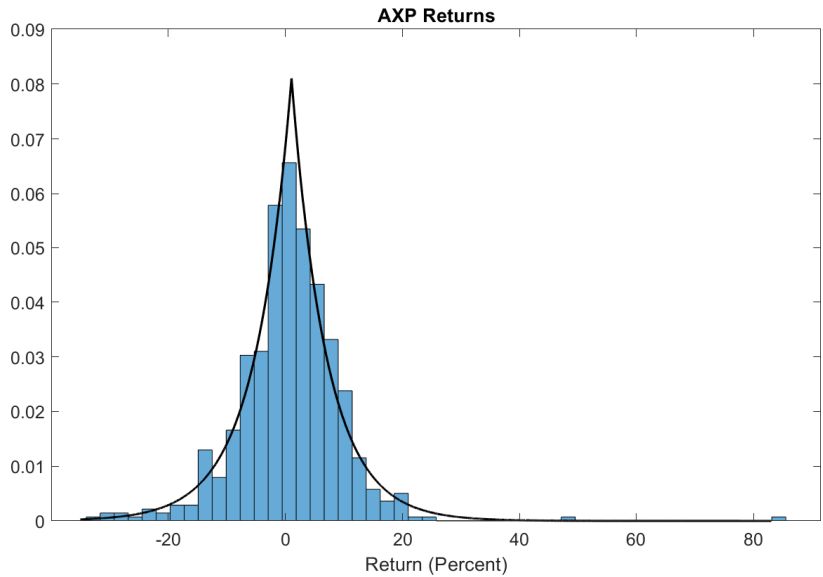}
	\end{center}
	\caption{Histogram of $n=580$ AXP monthly excess returns between 07/1972 and 10/2020 in percent. The fitted asymmetric Laplace density used for the 
		marginal density $f_{Y_i}(y_i)=g(y_i)$ is plotted as a black line.}\label{fig:5factFy}
\end{figure}

Linear regression is widely used
to estimate financial asset pricing models, where the dependent variable is
the (excess) return on a stock.
Yet stock returns are distributed far from Gaussian, so that a Gaussian regression model is 
mis-specified. 
To illustrate the regression copula model it is used to model
monthly excess returns on American Express Company (which has NYSE ticker symbol ``AXP'') using data from 07/1972 to 10/2020. 
The marginal distribution is assumed invariant with respect to observation, 
so that $F_{Y_i}(y_i)=G(y_i)$. A three parameter asymmetric Laplace distribution is
fit,
which better accounts for the distribution of returns as highlighted 
by~\cite{chen2012} and~\cite{taylor2019}. Figure~\ref{fig:5factFy} plots the 
density of the fitted margin, which is both asymmetric and has very heavy tails.

The monthly values of the
five factors suggested and described by~\cite{fama2015five} were used as covariates.
These are market risk (MktRf), size (SMB), value (HML), profit (RMW) and investment (CMA)
factors, with data
obtained from Kenneth French's website. The first three factors are widely 
employed, while the inclusion of the additional two factors RMW and CMA is more controversial. 
The linear regression copula
constructed using the horseshoe prior for $\betavec$ was estimated using Algorithm~\ref{alg:regcop}. Even though the implicit copula is 
of dimension $n=580$, employing the conditional likelihood at~\eqref{eq:regclike}
means that estimation is tractable, with a computation time of only 
32s to draw 10,000 iterates on a standard laptop.

\begin{table}[tbh]
	\caption{Posterior estimates of the regression copula model parameters for the
		American Express Company five factor asset pricing regression}\label{tab:5fact}
	\begin{center}
		\begin{tabular}{lccccc}\hline \hline
			Label &\multicolumn{5}{c}{Covariate}\\ \cline{2-6}
			& MktRf & SMB &HML &RMW &CMA \\
		 $\widehat{\betavec}$ &0.1889 &-0.0351 &0.0441 &-0.0020 &-0.0303\\
		95\% Interval &\footnotesize{(0.163,0.215)} &\footnotesize{(-0.067,-0.003)} &\footnotesize{(0.001,0.085)} &\footnotesize{(-0.031,0.024)} &\footnotesize{(-0.092,0.029)} \\
		$\widehat{\lambdavec}$ &0.0632 &0.0316 &0.0425 &0.0203 &0.1493\\
		MH Acceptance Rate &85\% &84\% &84\% &78\% &85\% \\
		\hline\hline
		\end{tabular}
	\end{center}
The first rows report
the posterior mean of $\betavec$ and the 95\% posterior probability intervals for each
covariate. The next rows report the posterior mean of $\lambdavec$, along with the Metropolis-Hasting (MH)
acceptance rate for each element. In addition, the posterior mean of $\tau$ is 0.0715 with an MH acceptance
rate of 92\%.  
\end{table}

Table~\ref{tab:5fact} summarizes the posterior estimates
of the coefficients $\betavec$ and the regularization parameters $\lambdavec$.
Of the five covariates, only the traditional three (MktRt, SMB and HML) were significant (``significant'' here refers to the whether, or not, zero falls into the 95\% posterior intervals for each coefficient $\beta_i$.) Thus, evidence 
for the inclusion of the two new factors RMW and CMA is weak. Also reported are the 
Metropolis-Hastings acceptance rates for the parameters. These are all high, suggesting Step~2 in Algorithm~\ref{alg:regcop} is effective.

\begin{figure}[tb]	
	\begin{center}
		\includegraphics[width=0.7\textwidth]{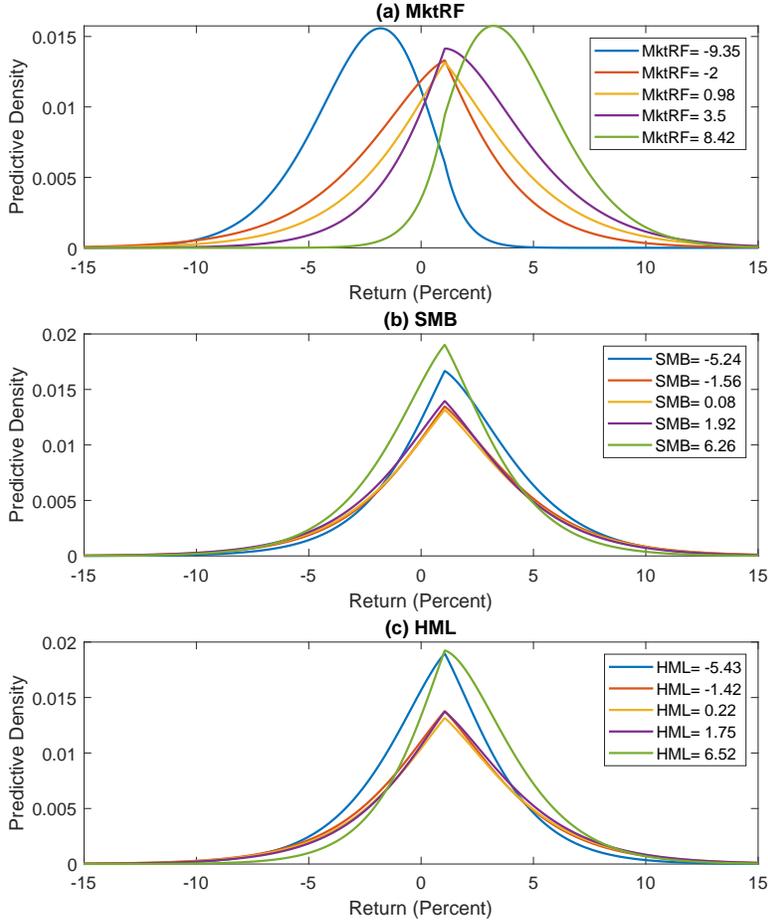}
	\end{center}
	\caption{Predictive densities of $Y$ (excess monthly return on AXP, in percent). Panel~(a) plots densities for five values of MktRf 
		corresponding to the 0.025, 0.25, 0.5, 0.75 and 0.975 observed quantiles, while
		setting the remaining covariates to their median values. Panels~(b) and~(c)
		repeat the process for covariates SMB and HML, respectively.}
		\label{fig:5factpred}
\end{figure}

To illustrate the effect 
of the three significant covariates on
the distribution of excess AXP returns, Figure~\ref{fig:5factpred} plots
the predictive density (estimated using $\hat{f}^{\mbox{\tiny Bayes}}_{\mbox{\tiny pred}}$) for different values of each covariate, 
setting the 
other four covariates equal to their median values. For example, in panel~(a) which 
focuses on variation in MktRf (the excess market return), 
the distribution is very different in location, spread, and shape for a typical 
month (MktRf=0.98) in comparison to a poor month ($\mbox{MktRf}=-9.35$)
or a strong month ($\mbox{MktRf}=8.42$). This highlights that the 
regression copula process combined with $G$ defines a 
distributional regression model, where each covariate affects the entire distribution
of $Y$.

\subsection{Further reading}
\subsubsection*{Extensions} 
While there are only $p=5$ covariates in the example here,
the regularization provided by the 
horseshoe prior allows $\xvec_i$ to be of much higher dimension $p$.
In particular, \cite{KleSmi2019} suggest forming $\xvec_i$ using a large 
number of functional basis terms, such as radial or p-spline bases. 
This produces a semiparametric distributional
regression model that these authors call a ``copula smoother''. 
\cite{KleNotSmi2019} instead suggest using the large number of terms 
from the output layer of a deep neural network (DNN) to form $\xvec_i$. The result is
a ``deep distributional regression'' method that combines the flexibility of a 
DNN with the probabilistic calibration of the copula model.

The implicit copula 
process described in Section~\ref{sec:lrc} can also be extended in several directions. \cite{klein2020bvs} derive the implicit copula for a linear regression with spike-and-slab priors for $\betavec$. This extends
popular Bayesian variable selection
methods to a dependent variable with an arbitrary marginal distribution.
\cite{smith+k19} extend the
homoscedastic regression for the auxiliary response $\tilde{Z}_i$
to a heteroscedastic regression.
The resulting implicit copula is a mixture of Gaussian copulas, and more flexible than the linear regression copula
outlined here. Implicit copula processes constructed from other regression models 
are also possible. 

\subsubsection*{Other approaches}
At~\eqref{eq:copregcdf} the marginals $F_{Y_j}$ are assumed independent
of the covariates, while the copula is not. 
In contrast, the copula can be assumed to be independent of the covariates, 
while the marginals are not; for examples, see~\cite{oakes2000}, \cite{pitt2006} and~\cite{song2009}. The first approach defines
a copula process for a univariate response, 
whereas the latter approach defines
a multivariate regression model for multiple response variables.

The implicit regression copulas outlined in Section~\ref{ssec:iregcopproc}
have a dependence structure that is a parametric function of the covariates
through the inversion of Sklar's theorem. For a linear regression, this is given by 
the expression for the correlation matrix $R(\xvec_{1:N},\thetavec)$ at~\eqref{eq:Rxtheta}. 
An alternative is to make either the parameters or dependence metrics of a copula $C$
smooth functions of the covariates without directly 
using inversion. Such models are called ``conditional copula models'',
and there is a extensive literature dealing with this case; for example, see~\cite{gijbels2011,veraverbeke2011,craiu2012,acar2013,sabeti2014,klein2016} and~\cite{vatter2018}.
Another approach is to treat the covariates as a random vector $\bm{X}$ and model it jointly 
with $Y$ in a copula model, from which the conditional distribution $Y|\bm{X}=\xvec$ can be derived.
This approach can be easily extended to multivariate responses, as in
\cite{zhao2019} who employ an elliptical copula with regularization of the 
parameter space provided by penalization of the coefficients of $\bm{X}$.

%% file: sec7.tex
\section{Discussion}\label{sec:discuss}
\subsection*{What is an implicit copula?}
While every copula $C$ has one or more implicit representations, 
it is often infeasible to derive the distribution $F_Z$ of the auxiliary variables. 
Instead, in this paper we consider implicit copulas to be those derived from a given
parametric continuous distribution $F_Z$. Knowledge of $F_Z$ makes estimation of these
implicit copulas tractable when
using likelihood-based estimation methods, such as Bayesian MCMC or variational
inference. This includes high-dimensional cases
where $C$ or $c$ cannot be computed in reasonable time, such
as the time series and regression copula processes discussed in this paper that
have dimension equal to the number of observations.

\subsection*{Comparison with vines}
Another copula family that can be
employed in high dimensions are vine copulas~\citep{joe1996,bedford2002}.
These are constructed from bivariate copula 
building blocks called ``pair-copulas'' by~\cite{AasCzaFriBak2009}.
By selecting different pair-copulas, vines can be constructed with 
a wide range of dependence structures; see~\cite{czado2019} for an overview. 
Vines are based on a decomposition
into conditional distributions. 
In some applications an appropriate decomposition
arises naturally, such as with time series as in~\cite{smith2010vine},
or when conditioning on latent factors as in~\cite{krupskii2013,krupskii2020}.
But, in general, there are many different possibilities~\citep{morales2010} and it can be difficult to select an appropriate choice, 
although there have been advances in 
approaches to do so~\citep{czado2019}. Another challenge in high dimensions is that 
it can be slow to evaluate the copula density and simulate from the vine, both of which
are necessary for parameter estimation and inference.
However, truncation as in~\cite{brechmann2012truncated}
or other simplifications, such as for stationary Markov time series~\citep{smith2010vine,smith2015,beare2015},
can alleviate these problems. In contrast, it is often unnecessary to evaluate
the implicit copula density when estimating the copula model, and simulation
from high dimensional
implicit copulas is typically fast and stable using Algorithm~\ref{alg:simicop}. 

\subsection*{Discrete and mixed marginals}
Copulas with discrete and mixed marginals are very increasingly popular~\citep{genest2007}, although parameter estimation is challenging.
Several approximate likelihood
approaches based on the continuous extension of discrete random variables studied by~\cite{denuit2005} have been suggested, although these typically exhibit significant bias in the copula parameter estimates; 
see~\cite{niko2013JSPI} and~\cite{nikoloulopoulos2016} for demonstrations using the Gaussian copula.
In contrast, Bayesian data augmentation approaches discussed here can evaluate
the posterior of such parametric copula models in high dimensions without 
resorting to approximating the likelihood.
For implicit copulas, estimation using data augmentation based on the extended likelihood
in Section~\ref{ssec:extlike} is popular in practice due to its simplicity and robustness. When using MCMC
sampling, as in~\cite{pitt2006} for the Gaussian copula, the posterior is evaluated exactly (up to Monte Carlo error) and can be
used in high dimensions as demonstrated by~\cite{danaher2011} and~\cite{dobra2011}. 
An alternative approach that can be used in even higher dimensions is variational inference,
as~\cite{loaiza2019VB} outline, although this is an approximate estimation method.

While the Gaussian copula
is by far the most popular choice when modeling the dependence of discrete data, it is not clear that it is always the best choice. 
For example, \cite{smith2012}
found that a skew $t$ copula provided a substantial improvement over a symmetric $t$ copula 
for 15-dimensional discrete data. While not explored here, implicit copula processes
can also be used for discrete time series data. Doing so provides an alternative to
the Markov vine copula models currently popular, as in~\cite{loaiza2019VB} and \cite{emura2021}.

\subsection*{Potential of implicit copula processes}
Finally, this article aims to highlight the
potential of time series and regression implicit 
copula processes. 
In machine learning they offer a computationally convenient avenue to extend existing deep models
to allow for uncertainty quantification. Examples include
\cite{salinas2019high} who do so using Gaussian copula processes
and~\cite{klein2020deep} who using the regression copulas in Section~\ref{sec:regcopprocess} with deep basis functions. 
The state space
copula proposed by~\cite{smithman2018} and outlined in Section~\ref{sec:sscop} also has substantial 
potential. Many existing statistical and econometric models can be
written in state space form, from simple time series models to smoothing splines. 
Combining their implicit copulas with flexible marginals
extends these models to more complex data distributions in a 
straightforward fashion.

%% file: append.tex
\appendix
\section{Evaluation of Marginals}\label{app:A}
\subsection*{Exact evaluation}
When estimating implicit copulas using likelihood-based methods, 
the marginal quantile functions $F_{Z_i}^{-1}$ and 
the densities $f_{Z_i}$ require evaluation. For more complex
implicit copulas each distribution function $F_{Z_i}(q)=\int_{-\infty}^{q} f_{Z_i}(s)\mbox{d}s$
is evaluated using univariate numerical integration. The quantile function can then be
obtained using a standard root finding algorithm such as Newton's method, which 
typically only requires a small number of steps to obtain an accurate value. This is much faster
than using Monte Carlo simulation from $F_{Z_i}$.

\subsection*{Fast interpolation}
However, for some applications the marginal quantile and density functions
of $Z_i$ have to be evaluated
at many observations. For example, this is the case with time series copulas
when $Z_i$ has a time invariant margin. 
To do so quickly the interpolation-based algorithm
in~\citet[App.~A]{smithman2018} can be used, which is fast to compute
once the interpolation is complete. These authors show it is
accurate for the state space models they study, while~\cite{yoshiba2018}
show the same algorithm is also accurate for the margin of a 
skew $t$ distribution. The algorithm is given below, and it
produces approximations for $\log(f_{Z_i})$ and $F^{-1}_{Z_i}$ far out into the tails
of the distribution, which can be used to evaluate the functions quickly at many values.

\begin{algorithm}[H]
\caption{\em (Interpolation of Quantile and Log-Density at $N$ Points)}
\label{alg:interpolate}
\begin{itemize}
\item[1.] Set $p_1=0.0001$ and $p_N=0.9999$, and evaluate both
$q_1=F_{Z_i}^{-1}(p_1)$ and $q_N=F_{Z_i}^{-1}(p_N)$
using a root finding algorithm (e.g. Newton's method).
\item[2.] Set step size to $\delta=(q_N-q_1)/(N-1)$, and a construct
uniform grid as $q_i=q_1+(i-1)\delta$, for $i=2,\ldots,N$; (e.g. $N=100$ is often sufficient).
\item[3.] For $i=1,\ldots,N$ (in parallel):
\begin{itemize}
 \item[3a.] Compute $p_i=F_{Z_i}(q_i)$ (possibly using univariate numerical integration)
 \item[3b.] Compute $b_i=\log(f_{Z_i}(q_i))$
\end{itemize}
\item[4.] Using an interpolation method (e.g. spline interpolation):
\begin{itemize}
 \item[4a.] Interpolate the points $\{(p_i,q_i);i=1,\ldots,N\}$ to obtain $F_{Z_i}^{-1}$
 \item[4b.] Interpolate the points $\{(q_i,b_i);i=1,\ldots,N\}$ to obtain $\log(p_{Z_i})$
\end{itemize}
\end{itemize}
\end{algorithm}

\section*{Acknowledgments} 
\noindent I would like to thank my co-authors on projects
involving implicit copulas, including Peter Danaher, Quan Gan, Richard Gerlach, Mohamad Khaled, Nadja Klein, Robert Kohn, Ruben Loaiza-Maya, Worapree Maneesoonthorn, David Nott
and Shaun Vahey. In particular, much of my work on time series copulas is joint
with Ruben Loaiza-Maya and Worapree Maneesoonthorn, and that on regression copulas is joint with Nadja Klein. I would also like to thank Professors
Ludger R\"{u}schendorf and Christian Genest for directing me to works in the copula literature, 
along with two referees, an associate editor and the editor Erricos Kontoghiorghes,  whose comments
have helped improve the manuscript. All errors are mine alone.
Last, I thank the organizers of the annual Computational and Financial Econometrics meetings, which
have provided me with an opportunity to present my work.